\documentclass[showpacs,twocolumn,preprintnumbers,amsmath,amssymb]{revtex4}
\usepackage{amsmath,amsfonts,latexsym,amssymb,graphicx,graphics,epsfig,subfigure,color,makeidx}
\usepackage{xcolor,diagbox}
\usepackage{multirow}
\usepackage[colorlinks,linkcolor=blue,anchorcolor=blue,citecolor=green,urlcolor=blue]{hyperref}
\usepackage{mathrsfs}
\usepackage{amsmath}

\begin{document}
\title{Characteristic modes of thick brane: resonances and quasinormal modes}

\author{Qin Tan$^{a}$$^{b}$$^{c}$\footnote{Qin Tan and Wen-Di Guo are co-first authors of the article.}}
\author{Wen-Di Guo$^{a}$$^{b}$$^{c}$}
\author{Yu-Peng Zhang$^{a}$$^{b}$$^{c}$}
\author{Yu-Xiao Liu$^{a}$$^{b}$$^{c}$\footnote{liuyx@lzu.edu.cn, corresponding author}}

\affiliation{
	$^{a}$Institute of Theoretical Physics $\&$ Research Center of Gravitation, Lanzhou University, Lanzhou 730000, China\\
	$^{b}$Key Laboratory of Quantum Theory and Applications of MoE, Lanzhou University, Lanzhou 730000, China\\
	$^{c}$Lanzhou Center for Theoretical Physics $\&$ Key Laboratory of Theoretical Physics of Gansu Province, Lanzhou University, Lanzhou 730000, China}
\begin{abstract}
In this work, we investigate the gravitational quasinormal modes (QNMs) and the gravitational resonances of a thick brane model. We use the asymptotic iteration and shooting methods to obtain the quasinormal frequencies (QNFs) of the brane. On the other hand, we investigate the resonances and their evolution numerically. The results show that the oscillations of the resonances equal (up to numerical error) to the real parts of the QNFs, while the damping rates of the resonances equal to the imaginary parts of the QNFs. The QNMs and resonances, both of them can be regarded as the characteristic modes of the thick brane, are closely related with each other. In addition, the lifetime of these QNMs could be very long, perhaps they might be detected in future accelerator or gravitational wave detector.

\end{abstract}
\pacs{ 04.50.-h, 11.27.+d}

\maketitle

\section{Introduction}
\label{Introduction}
In physics, characteristic modes are extremely important because they characterize the key features of a physical system. Investigating them could help us to understand a system. For a dissipative system, the characteristic modes are quasinormal modes (QNMs). For example, as the characteristic modes of a black hole, QNMs have attracted a lot of attention because they are expected to be able to carry information of a black hole~\cite{Berti:2009kk,Kokkotas:1999bd,Nollert:1999ji,Konoplya:2011qq,Cardoso:2016rao,Jusufi:2020odz,Cheung:2021bol}. QNMs also play a key role in other physical systems. It has been shown that there may exist a set of discrete modes in the thin and thick braneworld scenarios, i.e., QNMs of a brane~\cite{Seahra:2005wk,Seahra:2005iq,Tan:2022vfe}. Investigating them would help us to understand the properties of the branes.

Research on braneworld models has been ongoing for many years. Braneworld scenarios present a new viewpoint of spacetime and provide a new mechanism to solve the hierarchical problem between the Planck and Electroweak scales~\cite{ArkaniHamed:1998rs,Antoniadis:1998ig,Randall:1999ee}. One of the resolutions of the hierarchy problem is the Randall-Sundrum-I (RS-I) warped extra dimension model~\cite{Randall:1999ee}. It consists of two branes embedded in a five-dimensional anti-de-Sitter spacetime. The RS-I model was generalized to the RS-II one~\cite{Randall:1999vf} by pushing one brane to infinity. In the RS-II model, a remarkable character is that, even though the extra dimension is infinite, the four-dimensional Newtonian potential could be recovered. These brane models have been extensively investigated in various contexts such as black hole physics, particle physics, and cosmology~\cite{Shiromizu:1999wj,Tanaka:2002rb,Gregory:2008rf,Jaman:2018ucm,Adhikari:2020xcg,Geng:2020fxl,Geng:2021iyq,
Bhattacharya:2021jrn}. In the RS-II thin brane model, the thicknesses and inner structure of the brane are neglected, and so the energy density of the brane is a delta function along the extra dimension. However, when we concern the inner structure of a brane, we should investigate a thick brane generated by one or more matter fields. In fact, combining the domain wall model without gravity~\cite{Akama:1982jy,Rubakov:1983bb} and the RS-II model~\cite{Randall:1999vf}, DeWolfe et al proposed the thick brane models with gravity~\cite{DeWolfe:1999cp,Gremm:1999pj,Csaki:2000fc}. For further information on the development of the thick brane models, one can refer to the review articles~\cite{Liu:2017gcn,Ahluwalia:2022ttu,Dzhunushaliev:2009va}. Usually, the energy density of a thick brane is smooth. In previous literature, thick brane solutions in various gravity theories and localization of the gravitational zero mode and various matter fields on the branes were studied~\cite{Afonso:2007gc,Dzhunushaliev:2010fqo,Dzhunushaliev:2011mm,Geng:2015kvs,Melfo2006,Almeida2009,Zhao2010,
Chumbes2011,Liu2011,Bazeia:2013uva,Xie2017,Gu2017,ZhongYuan2017,ZhongYuan2017b,Zhou2018,Hendi:2020qkk,
Xie:2021ayr,Moreira:2021uod,Xu:2022ori,Silva:2022pfd,Xu:2022gth}. Besides the zero mode, there may exist massive Kaluza-Klein (KK) particles on the branes, which are particles beyond the standard model. If they are detected, it will open a new window to understand the nature of spacetime.

Recently, we investigated gravitational QNMs of a thick brane~\cite{Tan:2022vfe}. We found that there is a set of discrete QNMs in the thick brane. But the lifetimes of these QNMs are very short. In this paper, we aim to investigate whether there are long-lived QNMs in other thick branes, and if so, what are their properties. It has been known that there are long-lived massive modes called resonances in some thick brane models~\cite{Liu:2009ve,Almeida:2009jc,Cruz:2013uwa,Xu:2014jda,Csaki:2000pp,
Zhang:2016ksq,Sui:2020fty,Tan:2020sys,Chen:2020zzs,Moreira:2023pes,Belchior:2023gmr,Belchior:2023xgn}. Resonance is an important research topic in the investigation of thick brane. In most models, branes are dynamically generated by one or more background scalar fields. To be consistent with the standard model and the emergence of a four-dimensional Newtonian potential, it is crucial that zero modes of these matter fields and tensor fluctuations of gravity should be localized on these branes. Beyond zero modes, these models predict the existence of massive KK modes, representing novel particles. When considering a thick brane situated within a five-dimensional asymptotic anti-de Sitter spacetime, the effective potential experienced by KK modes traversing the extra dimension takes on a volcano-like shape. Consequently, the localization of massive KK modes onto the brane becomes unattainable. However, a finite number of these massive KK modes can be quasi-localized on the brane, recognized as resonant KK modes. In this paper, we focus on gravitational resonant KK modes, which notably contribute to the four-dimensional Newtonian potential. The spectrum of resonances can also reflect the structure of a brane. Moreover, the evolution of resonances is similar to that of the QNMs~\cite{Tan:2022uex}. We intuitively conjecture that there should be some relation between the resonances and the long-lived QNMs. To test this conjecture, we will investigate QNMs in the thick brane with inner structure.

The organization of the remaining part of this paper is as follows. In Sec.~\ref{BRANE WORLD MODEL}, we review the thick brane model and the linear metric tensor perturbation of the brane system. In Sec.~\ref{Quasi-normal modes of thick brane}, we solve the quasinormal frequencies (QNFs) of the thick brane by two semi-analytical methods. In Sec.~\ref{Resonances and QNMs}, we study the gravitational resonances of the thick brane and their evolution. And we compare the resonances and the QNMs of the brane. Finally, the conclusions and discussions are shown in Sec.~\ref{Conclusion}.

\section{Review of thick brane model}
\label{BRANE WORLD MODEL}
In this section, we will review the thick brane model in five-dimensional general relativity. Generally speaking, a thick brane could be generated by various matter fields~\cite{Xie2017,Gu2017,ZhongYuan2017,ZhongYuan2017b,Zhou2018,Chen:2020zzs,Hendi:2020qkk,
Xie:2021ayr,Moreira:2021uod,Xu:2022ori,Silva:2022pfd,Xu:2022gth}. The action is
\begin{eqnarray}
	S=\int d^5x\sqrt{-g}\left(\frac{1}{2\kappa^{2}_{5}}R-\mathcal{L}_{m} \right),\label{action}
\end{eqnarray}
where the five-dimensional gravitational constant $\kappa_{5}$ is set to $\kappa_{5}=1$ in this paper for convenience, and $\mathcal{L}_{m}$ is the Lagrangian of the matter fields. The dynamical equation is
\begin{eqnarray}
R_{MN}-\frac{1}{2}Rg_{MN}=T_{MN}.\label{field equation}
\end{eqnarray}
In this paper, capital Latin letters $M,N,\dots=0,1,2,3,5$ label the five-dimensional indices, Greek letters $\mu,\nu\dots=0,1,2,3$ label the four-dimensional ones, and Latin letters $i,j\dots=1,2,3$ label the three-dimensional space ones. The metric of the static flat brane is given by~\cite{DeWolfe:1999cp,Gremm:1999pj,Csaki:2000fc}
\begin{equation}
	ds^2=e^{2A(y)}\eta_{\mu\nu}dx^\mu dx^\nu+dy^2,
	\label{metric}
\end{equation}
where $e^{2A(y)}$ is the warp factor and $\eta_{\mu\nu}=\text{diag}(-1,1,1,1)$ is the four-dimensional Minkowski metric. The linear transverse-traceless tensor perturbation of the metric is given by
\begin{eqnarray}
	g_{MN}=\left(
	\begin{array}{cc}
		e^{2A(y)}(\eta_{\mu\nu}+h_{\mu\nu}) & 0\\
		0 & 1\\
	\end{array}
	\right)\label{perturbed metric}.
\end{eqnarray}
Here, $h_{\mu\nu}$ satisfies the transverse-traceless conditions
\begin{eqnarray}
	\partial_{\mu}h^{\mu\nu}=0=\eta^{\mu\nu}h_{\mu\nu}.\label{TTguage}
\end{eqnarray}
Combining the perturbed metric \eqref{perturbed metric} and the field equation \eqref{field equation}, the linear equation of the tensor perturbation is given by
\begin{eqnarray}
	\left(e^{-2A}\Box^{(4)}h_{\mu\nu}+h''_{\mu\nu}+4A'h'_{\mu\nu}\right)=0, \label{mainequation}
\end{eqnarray}
where $\Box^{(4)}=\eta^{\alpha\beta}\partial_{\alpha}\partial_{\beta}$. Transforming to conformally flat coordinates could simplify this wave equation. To this end we introduce the coordinate transformation $dz=e^{-A}dy$ and obtain the following metric
\begin{equation}
	ds^2=e^{2A(z)}(\eta_{\mu\nu}dx^\mu dx^\nu+dz^2).
	\label{conformalmetric}
\end{equation}
Now, the wave equation~(\ref{mainequation}) reads
\begin{equation}	\left[\partial^{2}_{z}+3(\partial_{z}A)\partial_{z}+\Box^{(4)}\right]h_{\mu\nu}=0.\label{conformalequation1}
\end{equation}
Making the following ansatz~\cite{Seahra:2005iq}
\begin{equation}
	h_{\mu\nu}=e^{-\frac{3}{2}A(z)}\Phi(t,z)e^{-i a_{j}x^{j}}\epsilon_{\mu\nu}, ~~~~\epsilon_{\mu\nu}=\text{constant},\label{decomposition1}
\end{equation}
we can rewrite Eq.~(\ref{mainequation}) as
\begin{equation}
	-\partial_{t}^{2}\Phi+\partial_{z}^{2}\Phi-U(z)\Phi-a^{2}\Phi=0,\label{evolutionequation}
\end{equation}
where
\begin{eqnarray}
	U(z)=\frac{3}{2}\partial_{z}^{2}A+\frac{9}{4}(\partial_{z}A)^{2} \label{effectivepotential}
\end{eqnarray}
is the effective potential and the parameter $a=\sqrt{\delta^{ij}a_i a_j}$ is constant of separation of variables. Further decomposing the function $\Phi(t,z)$ as
\begin{eqnarray}
	\Phi(t,z)=e^{-i\omega t}\phi(z),\label{decomposition2}
\end{eqnarray}
we can obtain a Schr\"odinger-like equation
\begin{equation}
	-\partial_{z}^{2}\phi(z)+U(z)\phi(z)=m^{2}\phi(z),\label{Schrodingerlikeequation}
\end{equation}
where
\begin{equation}
m^2=\omega^2-a^2\label{momega}
\end{equation}
is the mass of the KK mode. The Schr\"odinger-like equation \eqref{Schrodingerlikeequation} can be factorized as
\begin{equation}
	QQ^{\dagger}\phi(z)=m^{2}\phi(z), \label{SchrodingerEq2}
\end{equation}
where $Q$ and $Q^{\dagger}$ are defined as
\begin{equation}
	Q=\partial_{z}+\frac{3}{2}\partial_{z}A	,~~~~~~~~Q^{\dagger}=-\partial_{z}+\frac{3}{2}\partial_{z}A.
\end{equation}
Then, the dual equation corresponding to Eq. (\ref{SchrodingerEq2}) can be obtained as
\begin{equation}
Q^{\dagger}Q\tilde{\phi}(z)
=\left(-\partial_{z}^{2}+U^{\text{dual}}(z)\right)\tilde{\phi}(z)
=m^{2}\tilde{\phi}(z),\label{dualSchrodingerlikeequation0}
\end{equation}
where the dual potential $U^{\text{dual}}(z)$ of the effective potential~\eqref{effectivepotential} is given by
\begin{eqnarray}	
U^{\text{dual}}(z)=
-\frac{3}{2}\partial_{z}^{2}A+\frac{9}{4}(\partial_{z}A)^{2}.\label{dualeffpotentialform}
\end{eqnarray}
According to the super-symmetric quantum mechanics, the effective potential and the dual potential will share the same spectrum of massive excited states~\cite{Cooper:1994eh,Ge:2018vjq}. This property greatly facilitates the calculation of QNMs of a thick brane.

\section{Quasinormal modes of thick brane}
\label{Quasi-normal modes of thick brane}
\begin{figure}
	\subfigure[~The effective potential~(\ref{epform})]{\label{figeffective1}
		\includegraphics[width=0.22\textwidth]{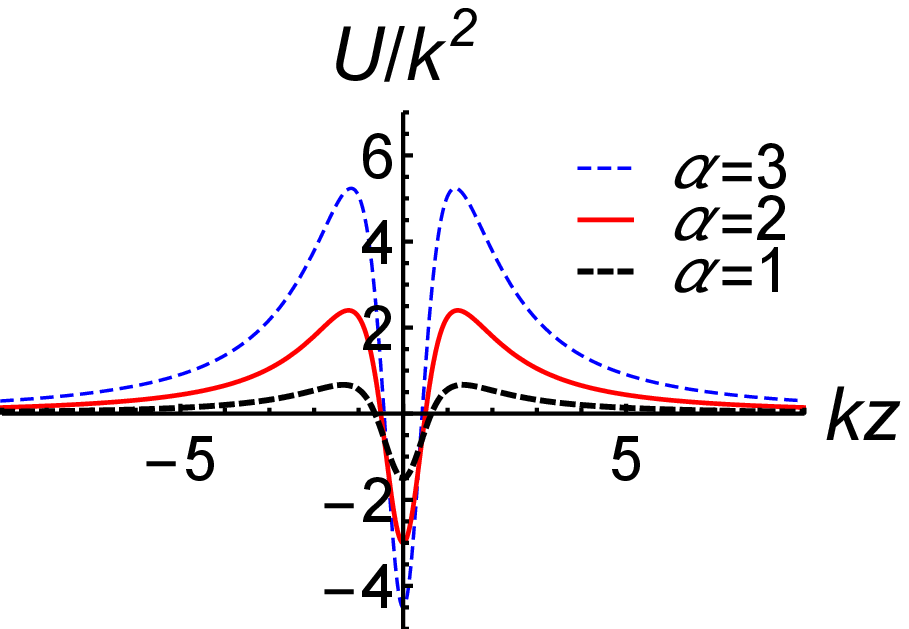}}
	\subfigure[~The dual effective potential~~\eqref{depform}]{\label{figeffective2}
		\includegraphics[width=0.22\textwidth]{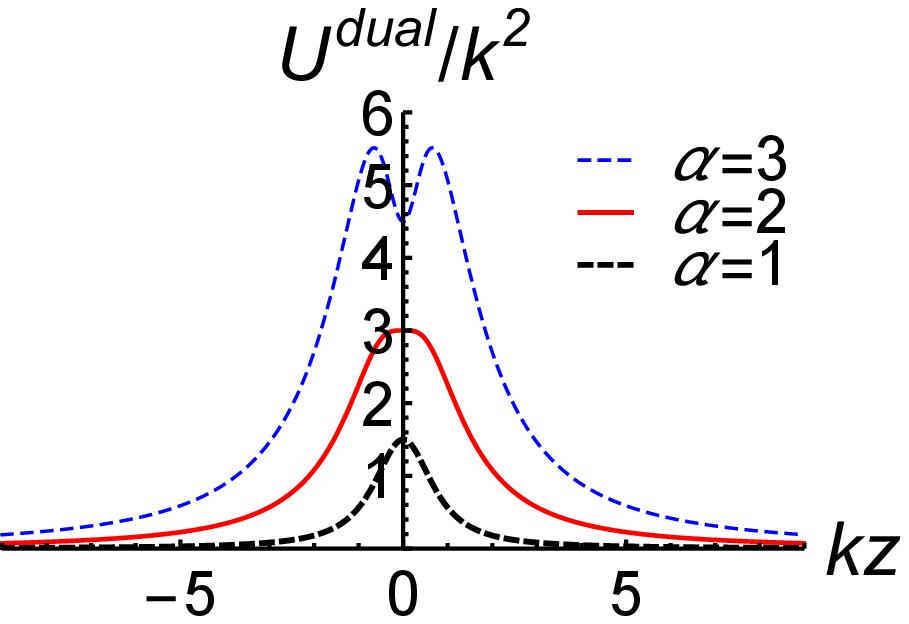}}	
	\subfigure[~The effective potential~(\ref{epform})]{\label{figeffective3}
		\includegraphics[width=0.22\textwidth]{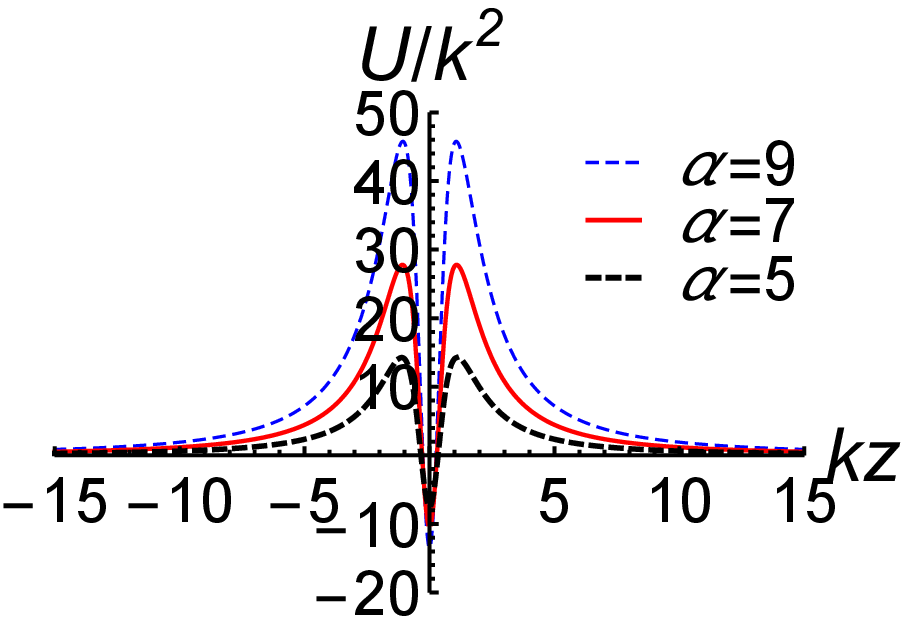}}
	\subfigure[~The dual effective potential~~\eqref{depform}]{\label{figeffective4}
		\includegraphics[width=0.22\textwidth]{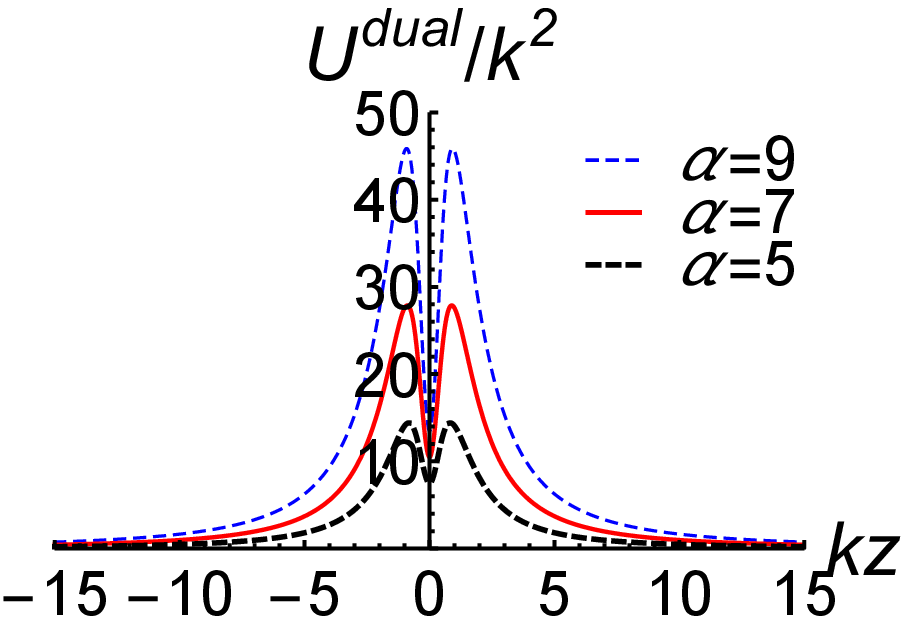}}
	\caption{Plots of the effective potential~\eqref{epform} and the dual effective potential~\eqref{depform}.}\label{dUandU}
\end{figure}

In this section, we investigate the QNMs of the thick brane. Since the Schr\"odinger-like equation \eqref{Schrodingerlikeequation} was obtained under the conformally flat metric \eqref{conformalequation1}, we focus on the conformally flat coordinate $z$ to study the QNMs of the thick brane. We choose the following warp factor~\cite{Csaki:2000fc}
\begin{equation}
A(z)=-\frac{\alpha}{2}\ln\left( k^{2}z^{2}+1 \right),\label{warpfactorz}
\end{equation}	
where $k$ has mass dimension one. The parameter $\alpha$ is a dimensionless constant and $\alpha>\frac{1}{3}$ to ensure the zero mode of gravity can be bound on the brane~\cite{Csaki:2000fc}. Substituting the warp factor~\eqref{warpfactorz} into the effective potential~\eqref{effectivepotential} and dual potential~\eqref{dualeffpotentialform}, we obtain the specific forms of the effective potential and the dual potential
\begin{eqnarray}
U(z)&=&\frac{3 \alpha  k^2 \left((3 \alpha +2) k^2 z^2-2\right)}{4 \left(k^2 z^2+1\right)^2},\label{epform}\\
U^{\text{dual}}(z)&=&\frac{3 \alpha  k^2 \left((3 \alpha -2) k^2 z^2+2\right)}{4 \left(k^2 z^2+1\right)^2}\label{depform}.
\end{eqnarray}
Plots of the above two potentials are shown in Fig.~\ref{dUandU}. It can be seen that, the heights of the effective potential and the dual potential increase with the parameter $\alpha$. As $\alpha$ increases, a quasi-well appears in the dual potential, which generally implies that there might be gravitational resonances. Next, we use the asymptotic iteration method (AIM)~\cite{Ciftci:2003As,ciftci:2005co,Cho:2011sf} and the shooting method~\cite{Pani:2013pma} to solve the QNFs of the thick brane. In the process of using AIM to solve the QNFs, we find that it is more convenient to use the dual potential than the effective potential to solve the QNFs of the thick brane. Moreover, the spectra of the QNMs for the two potentials are the same~\cite{Ge:2018vjq}. Therefore, the dual potential is used in this work to solve the QNFs of the brane.

First, we shall briefly review the idea behind the AIM~\cite{Ciftci:2003As,ciftci:2005co}. The AIM is an analytical and approximate method proposed by Ciftci et al. for solving second-order linear differential equations, especially the eigenvalue problems that often appear in theoretical and mathematical physics. Many eigenvalue problems in relativistic and non-relativistic quantum mechanics can be solved using the AIM~\cite{Ciftci:2005xn,Bayrak:2006qt,Bayrak:2006wu,Champion:2008hg}. Since solving QNMs is also an eigenvalue problem, the AIM can also be used. For a second-order linear differential equation of the form
\begin{equation}
y''(x)=\lambda_{0}(x)y'(x)+s_{0}(x)y(x),\label{2orderdiffeq}
\end{equation}
where $\lambda_{0}(x)$ and $s_{0}(x)$ are $C^{\infty}$ functions with $\lambda_{0}(x)\neq0$. Differentiating Eq.~\eqref{2orderdiffeq} with respect to $x$, we can obtain
\begin{equation}
	y'''(x)=\lambda_{1}(x)y'(x)+s_{1}(x)y(x),
\end{equation}
where
\begin{eqnarray}
	\lambda_{1}(x)&=&\lambda'_{0}+s_{0}+\lambda_{0}^{2},\\
	s_{1}(x)&=&s'_{0}+s_{0}\lambda_{0}.
\end{eqnarray}
In the AIM, one can use the invariant structure of the right-hand side of Eq.~\eqref{2orderdiffeq} to find a general solution. In fact, the $(n-1)$-th and $n$-th differentiations of Eq.~\eqref{2orderdiffeq} yield
\begin{eqnarray}
	y^{n+1}(x)&=&\lambda_{n-1}(x)y'(x)+s_{n-1}(x)y(x),\\
	y^{n+2}(x)&=&\lambda_{n}(x)y'(x)+s_{n}(x)y(x),	
\end{eqnarray}
where
\begin{eqnarray}
	\lambda_{n}(x)&=&\lambda'_{n-1}+s_{n-1}+\lambda_{0}\lambda_{n-1}, \label{AIMrelation1}\\ s_{n}(x)&=&s'_{n-1}+s_{0}\lambda_{n-1}\label{AIMrelation2}.
\end{eqnarray}
For sufficiently large $n$, the asymptotic aspect is introduced:
\begin{eqnarray}
\frac{s_{n}(x)}{\lambda_{n}(x)}=\frac{s_{n-1}(x)}{\lambda_{n-1}(x)}=\beta(x).\label{QNMscondition1}
\end{eqnarray}
The QNFs can be solved from the ``quantization condition"
\begin{eqnarray}
	s_{n}(x)\lambda_{n-1}(x)-s_{n-1}(x)\lambda_{n}(x)=0.\label{QNMscondition2}
\end{eqnarray}
But the above ``quantization condition" has an unappealing feature that for each iteration one must take the derivative of the $s(x)$ and $\lambda(x)$ terms of the previous iteration. This is inconvenient for numerical operations. Cho $et~al.$~\cite{Cho:2011sf} developed an improved version of the AIM which greatly improves the speed and accuracy of numerical calculation. The basic idea of the improved AIM is to expand $s(x)$ and $\lambda(x)$ using Taylor series at a point $\chi$
\begin{eqnarray}
	\lambda_{n}(x)&=&\sum_{i=0}^{\infty}c_{n}^{i}(x-\chi)^{i},\\
	s_{n}(x)&=&\sum_{i=0}^{\infty}d_{n}^{i}(x-\chi)^{i},
\end{eqnarray}
where $c_{n}^{i}$ and $d_{n}^{i}$ are the $i$-th Taylor coefficients of $\lambda_{n}$ and $s_{n}$, respectively. Expressions of $c_{n}^{i}$ and $d_{n}^{i}$ are
\begin{eqnarray}
c_{n}^{i}&=&(i+1)c_{n-1}^{i+1}+d^{i}_{n-1}+\sum_{k=0}^{i}c_{0}^{k}c_{n-1}^{i-k},\\
d_{n}^{i}&=&(i+1)d_{n-1}^{i+1}+\sum_{k=0}^{i}d_{0}^{k}c_{n-1}^{i-k}.
\end{eqnarray}
Thus, the ``quantization condition" ~\eqref{QNMscondition2} becomes
\begin{equation}
	d_{n}^{0}c_{n-1}^{0}-d_{n-1}^{0}c_{n}^{0}=0\label{QNMscondition3}.
\end{equation}
Now we have a set of recursion relations, which do not require derivative operators.

\begin{figure}
	\subfigure[The first QNF]{\label{figarem1}
		\includegraphics[width=0.22\textwidth]{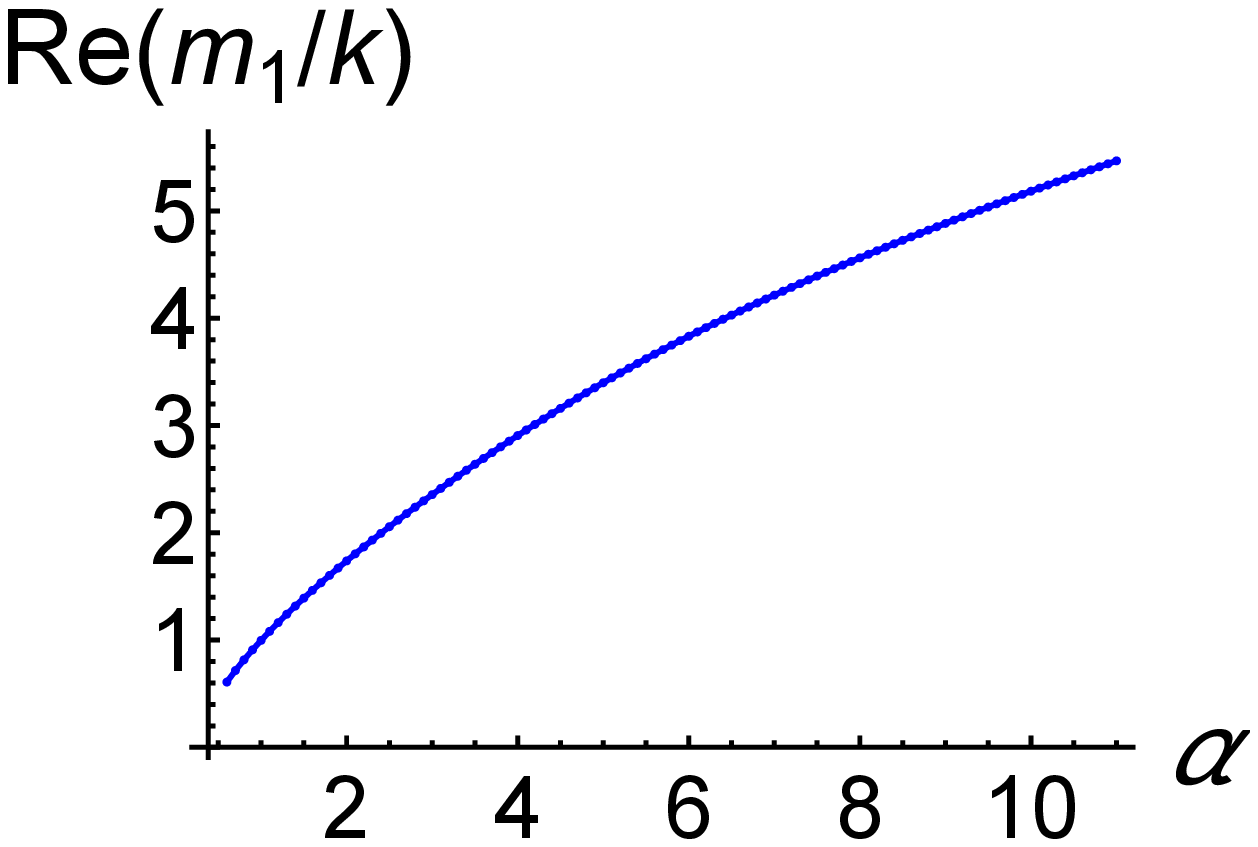}}
	\subfigure[The first QNF]{\label{figaimm1}
		\includegraphics[width=0.22\textwidth]{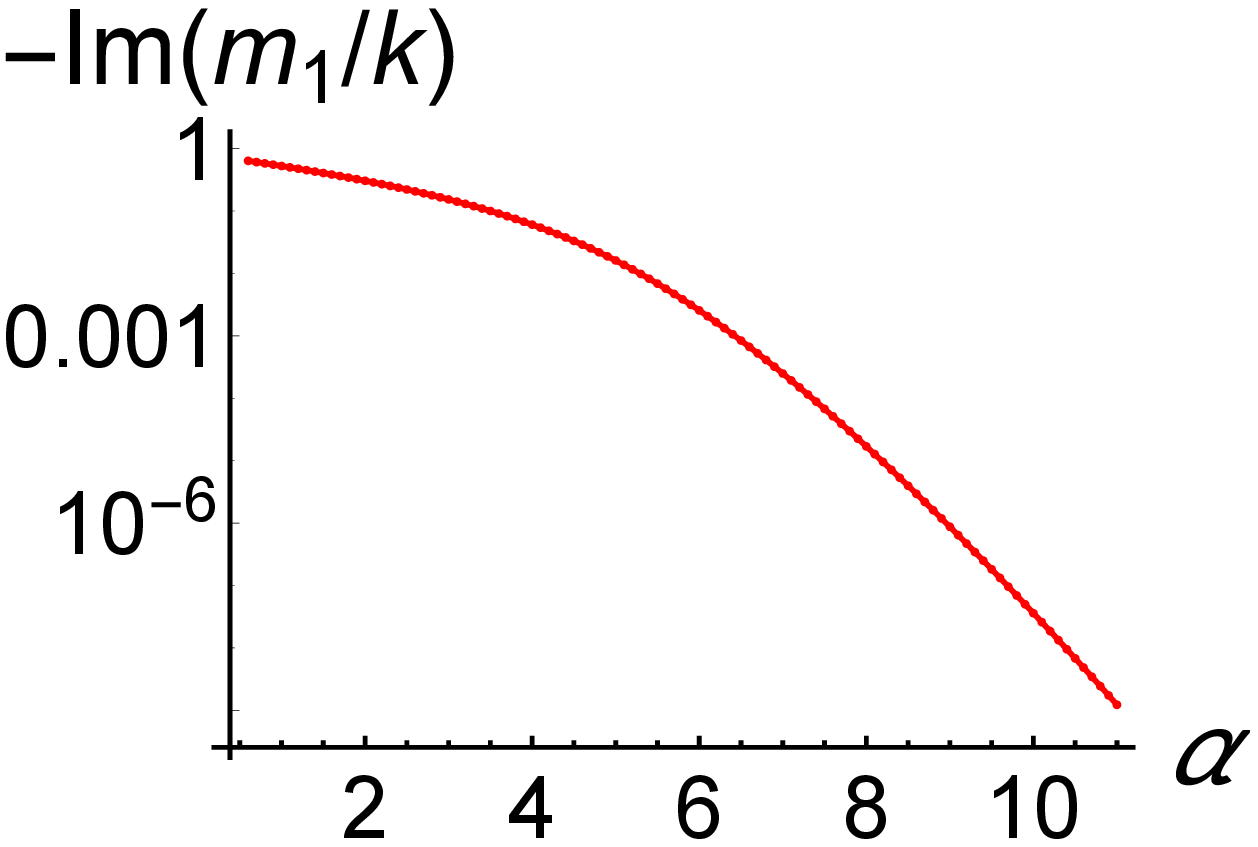}}
	\subfigure[The second QNF]{\label{figarem2}
	\includegraphics[width=0.22\textwidth]{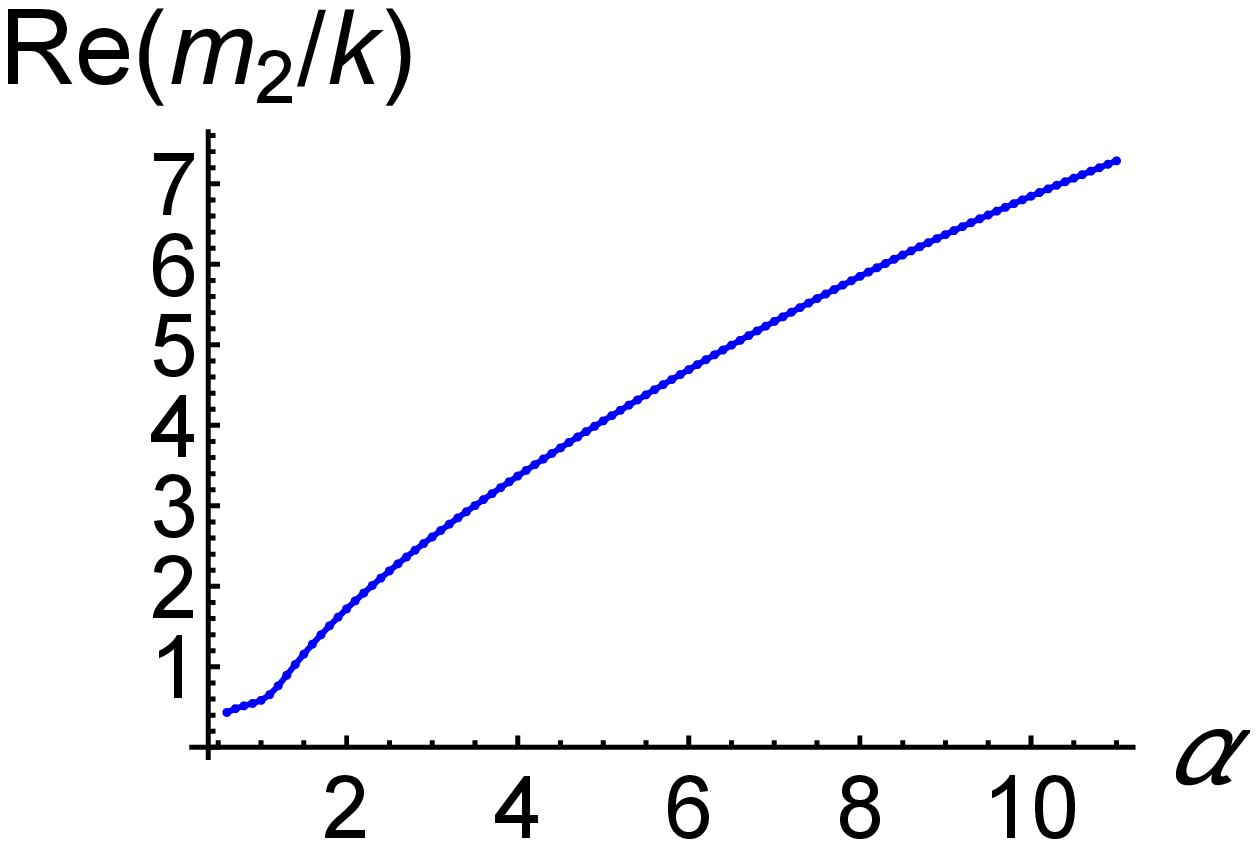}}
\subfigure[The second QNF]{\label{figaimm2}
	\includegraphics[width=0.22\textwidth]{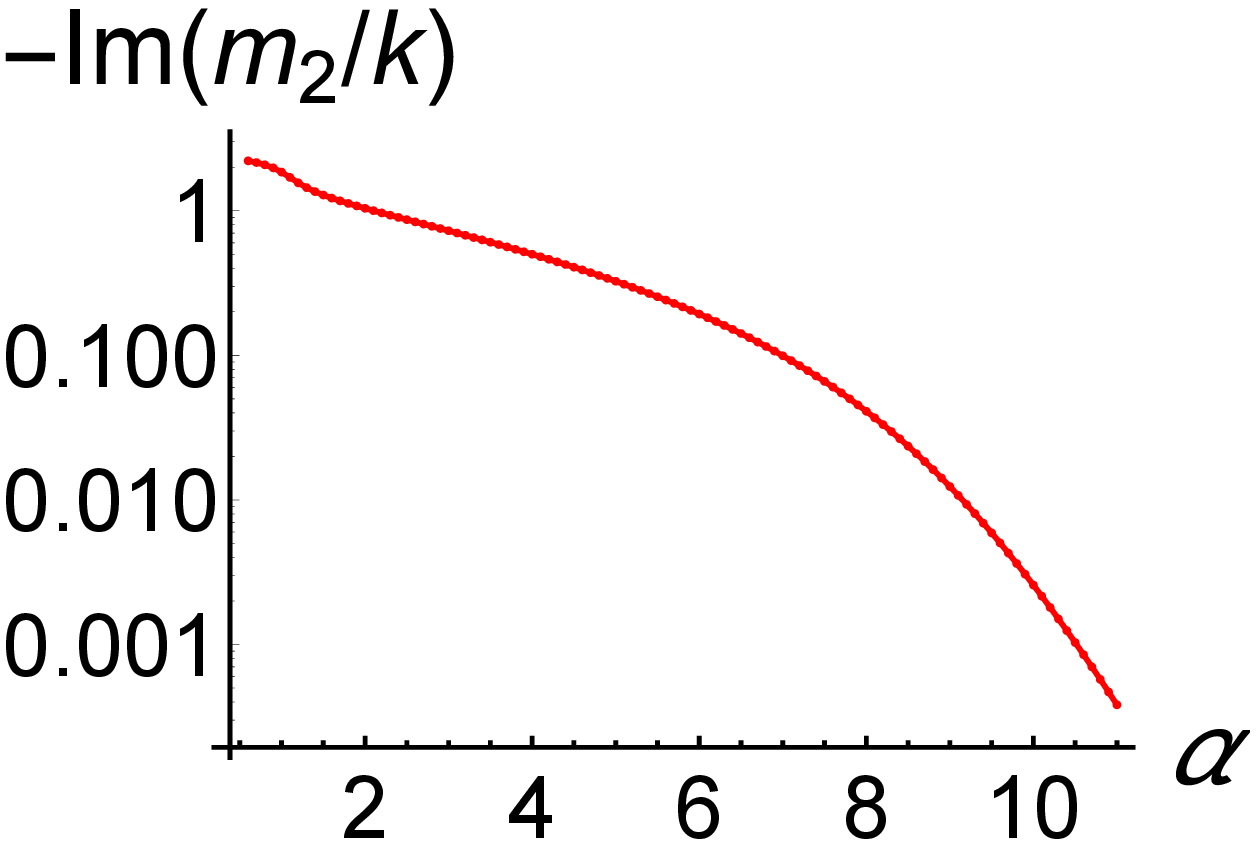}}
	\subfigure[The third QNF]{\label{figarem3}
	\includegraphics[width=0.22\textwidth]{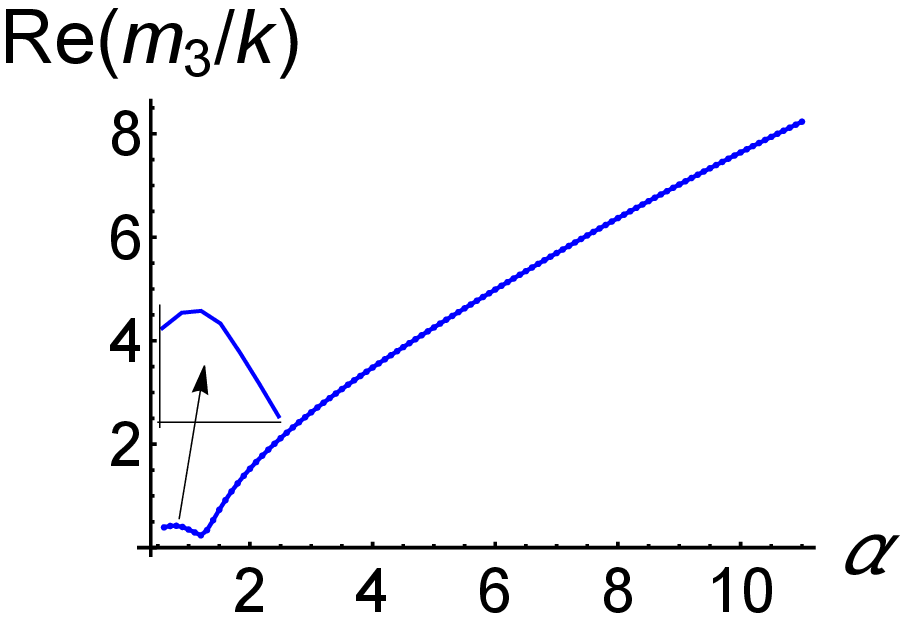}}
\subfigure[The third QNF]{\label{figaimm3}
	\includegraphics[width=0.22\textwidth]{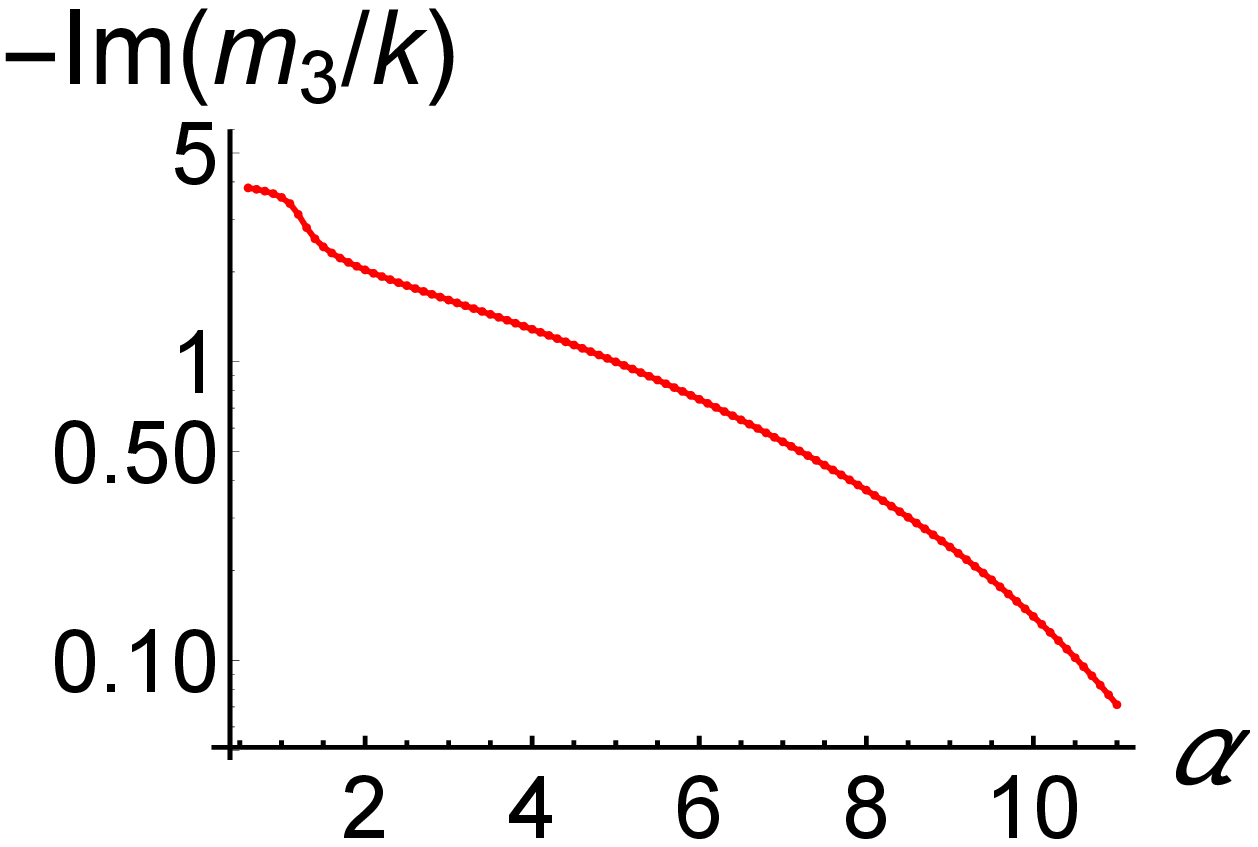}}
	\caption{The relation between the real parts (left panel) and imaginary parts (right panel) of the first three quasinormal frequencies and the parameter $\alpha$. Note that we use a logarithmic scale for the imaginary parts of the quasinormal frequencies.}\label{alpham}
\end{figure}

With the choice of the warp factor~\eqref{warpfactorz}, the Schr\"odinger-like equation~\eqref{dualSchrodingerlikeequation0} is
\begin{eqnarray}
	-\partial_{z}^{2}\tilde{\phi}(z)+\left(\frac{3 \alpha  k^2 \left((3 \alpha -2) k^2 z^2+2\right)}{4 \left(k^2 z^2+1\right)^2}-m^{2}\right)\tilde{\phi}(z)=0.\nonumber\\ \label{dualSchrodingerlikeequation}
\end{eqnarray}
For a massive KK mode, the thick brane is a dissipative system. Therefore, we should impose the maximally dissipative boundary condition. That is, the wave function should be purely outgoing at spatial infinity and purely ingoing at negative spatial infinity, such that
\begin{equation}
	\label{boundaryconditions}
	\tilde{\phi}(z)\sim\left\{
	\begin{aligned}
		e^{im z}, &~~~~~z\to\infty& \\
		e^{-im z},  &~~~~~z\to-\infty&.
	\end{aligned}
	\right.
\end{equation}
Note that Eq.~(\ref{dualSchrodingerlikeequation}) does not contain a first derivative term, which means that $\lambda_{0}$ must be zero. As a result, the asymptotic iteration method cannot be applied directly. To overcome this limitation, we need to perform a coordinate transformation to obtain an equation that includes a nonvanishing first derivative term. In addition, the AIM works better on a finite domain. Thus, we transform the coordinate $z$ to $u$ with
$u=\frac{\sqrt{4k^2 z^2+1}-1}{2k z}$. Then Eq.~\eqref{dualSchrodingerlikeequation} becomes
 \begin{eqnarray}
 &&\left(\frac{m^2}{k^2}-\frac{3\alpha \left(u^2-1\right)^2 \left(2 u^4-3 u^2(\alpha-2)+2\right)}{4\left(u^4-u^2+1\right)^2}\right) \tilde{\phi}(u)\nonumber\\
 &&+\frac{\left(u^2-1\right)^3 \left(\left(u^4-1\right) \tilde{\phi}''(u)+2 u \left(u^2+3\right) \tilde{\phi}'(u)\right)}{\left(u^2+1\right)^3}=0,\nonumber\\\label{dualSchrodingerlikeequation1}
 \end{eqnarray}
where $-1<u<1$. The boundary conditions~(\ref{boundaryconditions}) can be rewritten as
\begin{equation}
	\label{transformboundaryconditions}
	\tilde{\phi}(u)\sim\left\{
	\begin{aligned}
		e^{-\frac{i m/k }{2 u-2}}, &~~~ u\to 1& \\
		e^{\frac{i m/k }{2 u+2}}, &~~~ u\to -1&
	\end{aligned}.
	\right.
\end{equation}
Next we define a new function ${\psi}(u)$: 
\begin{eqnarray}
	\tilde{\phi}(u)=\psi (u) e^{-\frac{i m/k }{2 u-2}} e^{\frac{i m/k }{2 u+2}}.\label{boundarysolutions}
\end{eqnarray}
Substituting this expression into Eq.~\eqref{dualSchrodingerlikeequation1} we can obtain the equation for ${\psi}(u)$
\begin{equation}
\psi''(u)=\lambda_{0}(u)\psi'(x)+s_{0}(u)\psi(u),\label{2orderdiffeq_psi}
\end{equation}
where
\begin{eqnarray}
	\lambda_{0}(u)&=&-\frac{2 u \left(u^4+2 i \left(u^2+1\right) \frac{m}{k} +2 u^2-3\right)}{\left(u^2-1\right)^2 \left(u^2+1\right)},\label{lambda0}\\
	s_{0}(u)&=&\frac{1}{4 \left(u^2+1\right) \left(u^6-2 u^4+2 u^2-1\right)^2}\nonumber\\
	&\times& \Bigg[6 \alpha\left(u^{10}-5 u^6-5 u^4+1\right)+9 \alpha^2 u^2 \left(u^2+1\right)^3\nonumber\\
	&-&\frac{4m}{k} \left(u^4-u^2+1\right)^2 \left(u^2 (\frac{m}{k} -2 i)+\frac{m}{k} +2 i\right)\Bigg].\nonumber\\\label{s0}
\end{eqnarray}
Once the specific form of $\lambda_{0}$ and $s_{0}$ are obtained, the QNFs can be solved by performing the improved AIM. Note that in the improved AIM, we need to give a coordinate point $u_0$ to find the eigenvalue. The choice of coordinate point $u_0$ is arbitrary in principle, but in practice the choice of different points will affect the accuracy and speed of calculation~\cite{Cho:2011sf}. For obvious symmetry reasons we set $u_0=0$. We plot the effect of the parameter $\alpha$ on the real and imaginary parts of the QNFs for the thick brane in Fig.~\ref{alpham}. It can be seen that the real parts of the first two QNFs increase with $\alpha$, while the imaginary parts of the first three QNFs decrease with $\alpha$. This is because the height of the dual potential and the lowest point of the quasi-well increase with the parameter $\alpha$. However, the real part of the third QNF shows a different behavior: it first increases, then decreases (as can be seen from the sub-figure in Fig.~\ref{figarem3}), and then increases with $\alpha$. This may be related to the structure of the dual potential. Since the dual potential also has a double peak structure when the parameter $\alpha$ is large, the result of the AIM may not be accurate. Therefore, we solve the QNFs by the shooting method~\cite{Pani:2013pma} and compare them with the results of the AIM, which are shown in Tab.~\ref{tab1}. We find that the results of the two methods agree well with each other. This enhances the credibility of the results. Note that, since there is a bound zero mode on the thick brane, we denote the QNM with the longest lifetime as the first overtone, i.e., $n=1$. This is different from the case of a black hole system. On the other hand, since the imaginary part of a QNF relates to the lifetime of a KK mode, so for a large enough $\alpha$, there might be long-lived KK modes. In previous investigations, there are long-lived KK modes called resonances on some thick branes. We will  investigate the relation between the resonances and the long-lived QNMs in the next section.

\begin{table*}[htbp]
	\begin{tabular}{|c|c|c|c|}
		\hline
		$\;\;\alpha\;\;$  &
		$\;\;n\;\;$  &
		$\;\;\text{Asymptotic iteration method}\;\;$  &
		$\;\;\;\;\;\;\;\;\text{Shooting method}\;\;\;\;\;\;\;$ \\
		\hline
		~  &~   &~~~~$\text{Re}(m/k)$  ~~  $\text{Im}(m/k)~~$  &$~~~~~~\text{Re}(m/k)$ ~~ $\text{Im}(m/k)~~$       \\
		1  &1   &0.99702~~ -0.526362       &~~0.99702~~  -0.526365       \\
		2  &1   &1.73769~~ -0.305138          &~~1.73769~~ -0.305138  \\		
		3  &1   &2.35548~~ -0.153401          &~~2.35548~~ -0.153401   \\
		   &2   &2.61306~~ -0.727270          &~~2.61306~~ -0.727270   \\
		4  &1   &2.90598~~ -0.060511          &~~2.90598~~~-0.060511  \\
		   &2   &3.36982~~ -0.500264          &~~3.36982~~~-0.500264  \\
		5  &1   &3.39797~~ -0.016113          &~~3.39797~~  -0.016113  \\
		   &2   &4.05532~~ -0.325381          &~~4.05532~~ -0.325381  \\	
		6  &1   &3.83141~~ -0.002554          &~~3.83141~~  -0.002554   \\
		   &2   &4.69248~~ -0.193092          &~~4.69248~~ -0.193092   \\
		   &3   &4.98876~~ -0.748896          &~~4.98876~~ -0.748896   \\
		\hline
	\end{tabular}
	\caption{Low overtone modes using the AIM and shooting method.\label{tab1}}
\end{table*}

\section{Resonances and quasinormal modes}
\label{Resonances and QNMs}
\begin{figure}
	\subfigure[~the configuration of resonance]{\label{figresonancephi}
		\includegraphics[width=0.22\textwidth]{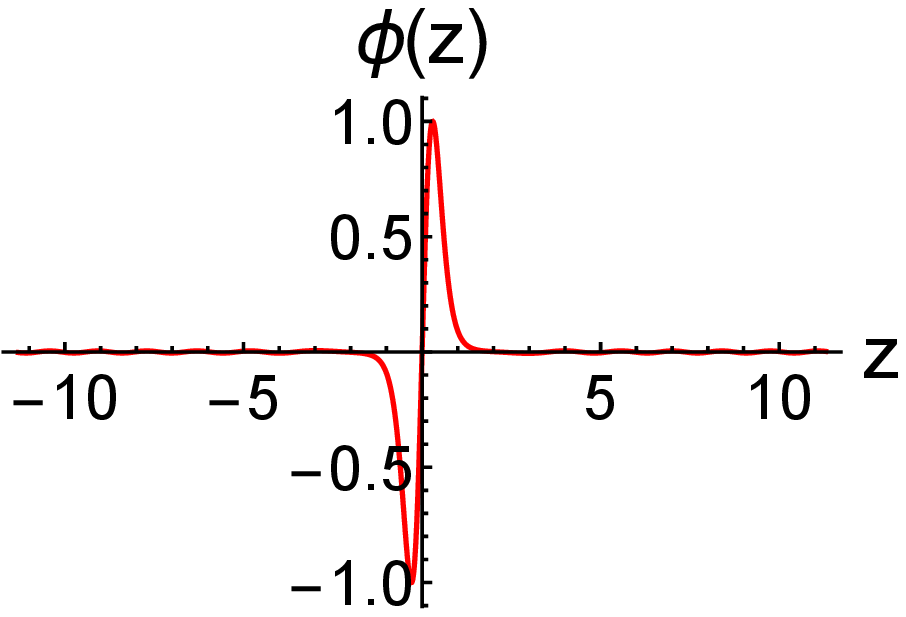}}
	\subfigure[~the configuration of nonresonance]{\label{fignonresonance}
		\includegraphics[width=0.22\textwidth]{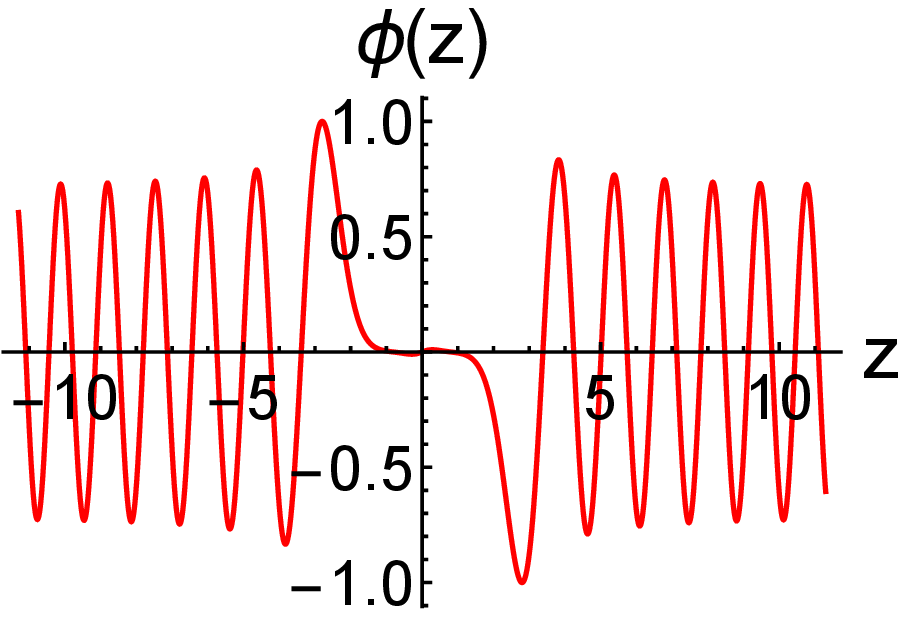}}	
	\caption{The configurations of the resonance and nonresonance.}\label{figphi}
\end{figure}

Resonance is an important research topic in the study of thick brane. In the previous investigations, resonances are regarded as a specific class of massive KK modes which could be quasi-localized on the branes~\cite{Liu:2009ve,Almeida:2009jc,Cruz:2013uwa,Xu:2014jda,Csaki:2000pp,
Zhang:2016ksq,Sui:2020fty,Tan:2020sys,Chen:2020zzs,Moreira:2023pes,Belchior:2023gmr,Belchior:2023xgn}. The resonance spectra vary with different brane configurations. In this study, we focus on the characterization of gravitational resonances for the thick brane model.  We use the relative probability method to identify all gravitational resonances. The relative probability is given by~\cite{Liu:2009ve}
\begin{eqnarray}
	P(m^{2})=\frac{\int^{z_{\text{b}}}_{-z_{\text{b}}}|\phi(z)|^{2}dz}
	{\int^{z_{\text{max}}}_{-z_{\text{max}}}|\phi(z)|^{2}dz},\label{relative probability}
\end{eqnarray}
where $\phi(z)$ is the solution of Eq.~\eqref{Schrodingerlikeequation}, $z_{\text{max}}=10z_{\text{b}}$ and $z_{\text{b}}$ is approximately the width of the brane. Figure~\ref{figphi} is a schematic diagram of resonance and nonresonance configurations. It can be observed that the amplitude of a resonance inside the quasi-well is significantly greater than the amplitude outside the well. Using the relative probability method, we can calculate the relative probability of a KK mode. If the relative probability $P(m^2)$ exhibits a peak around $m = m_n$ and this peak has a width at half maximum, it indicates the presence of a resonance with a mass $m_n$. Through this method, we can find all of the resonances.

\begin{figure}
	\subfigure[~$\alpha=5$]{\label{figPalpha5}
		\includegraphics[width=0.22\textwidth]{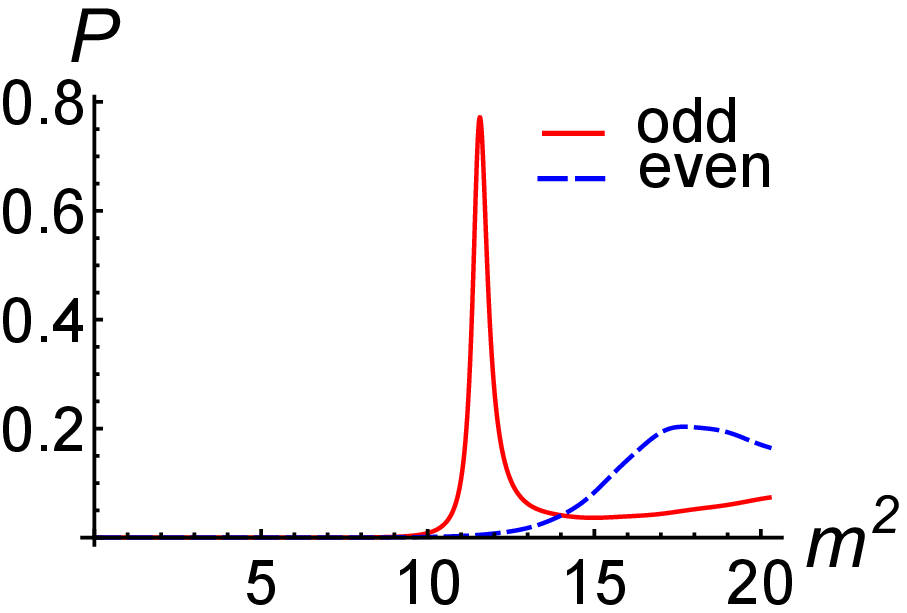}}
	\subfigure[~$\alpha=7$]{\label{figPalpha7}
		\includegraphics[width=0.22\textwidth]{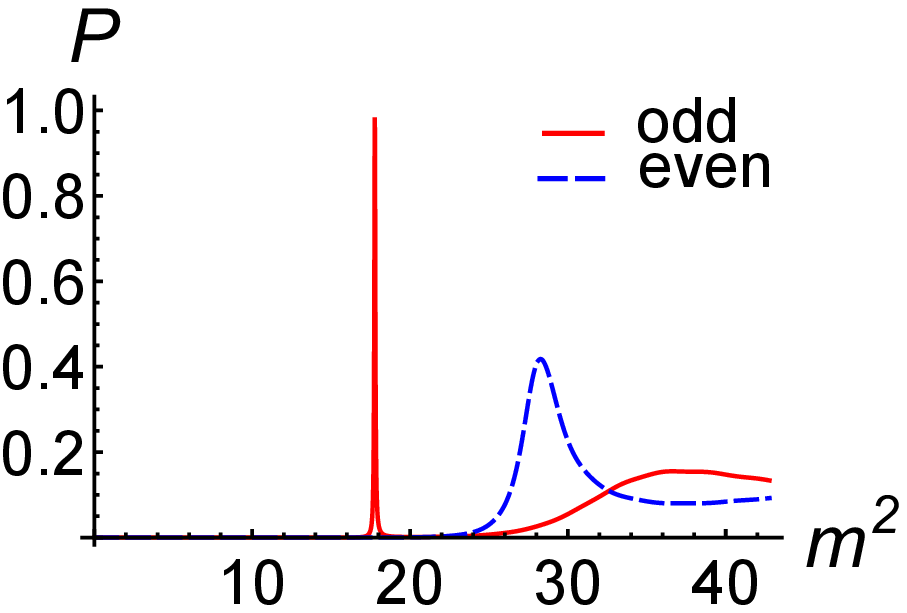}}	
	\subfigure[~$\alpha=9$]{\label{figPalpha9}
		\includegraphics[width=0.22\textwidth]{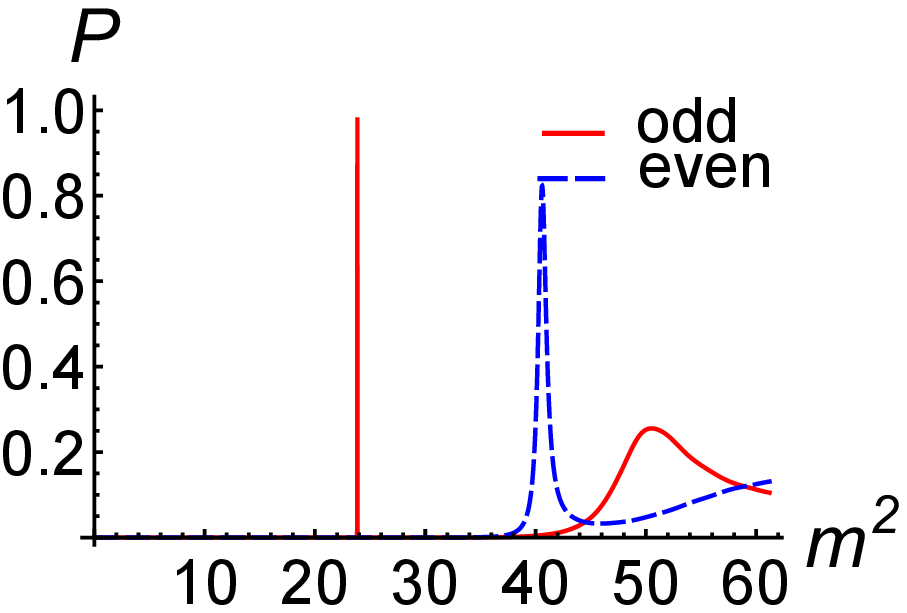}}	
	\caption{The influence of the parameter $\alpha$ on the relative probability $P(m^{2})$.}\label{figPalpha}
\end{figure}

\begin{figure*}
	\subfigure[~$\alpha=5$]{\label{figa5o1z1}
		\includegraphics[width=0.5\textwidth]{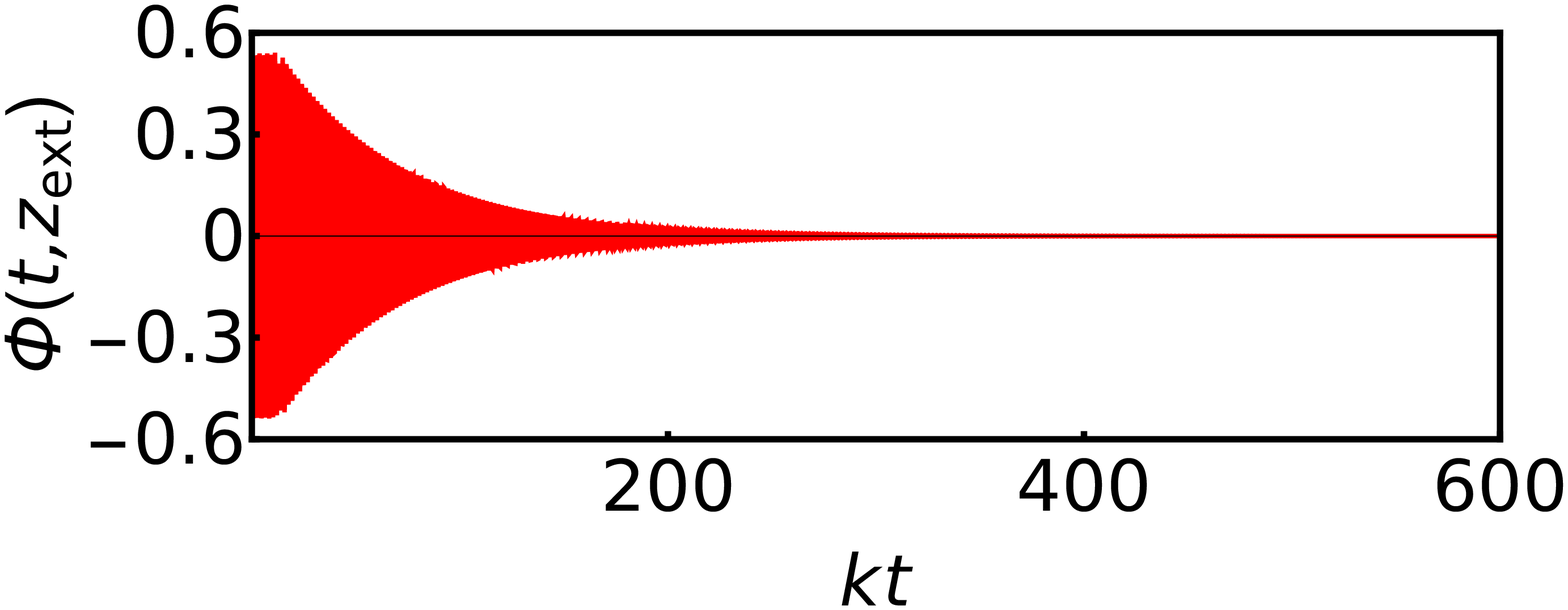}}
	\subfigure[~$\alpha=5$]{\label{figfouriera5o1z1}
		\includegraphics[width=0.35\textwidth]{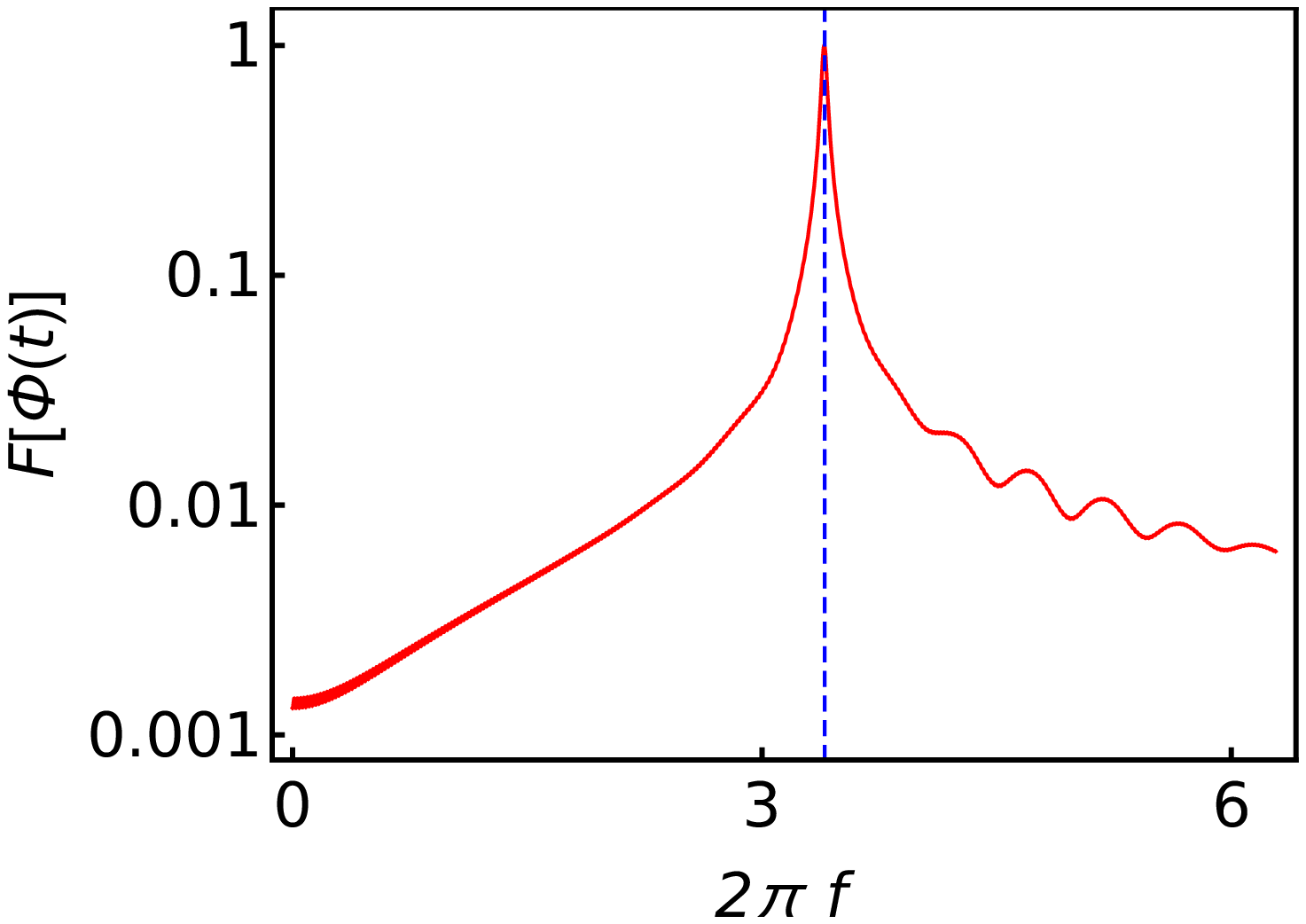}}
	
	\subfigure[~$\alpha=7$]{\label{figa7o1z1}
		\includegraphics[width=0.5\textwidth]{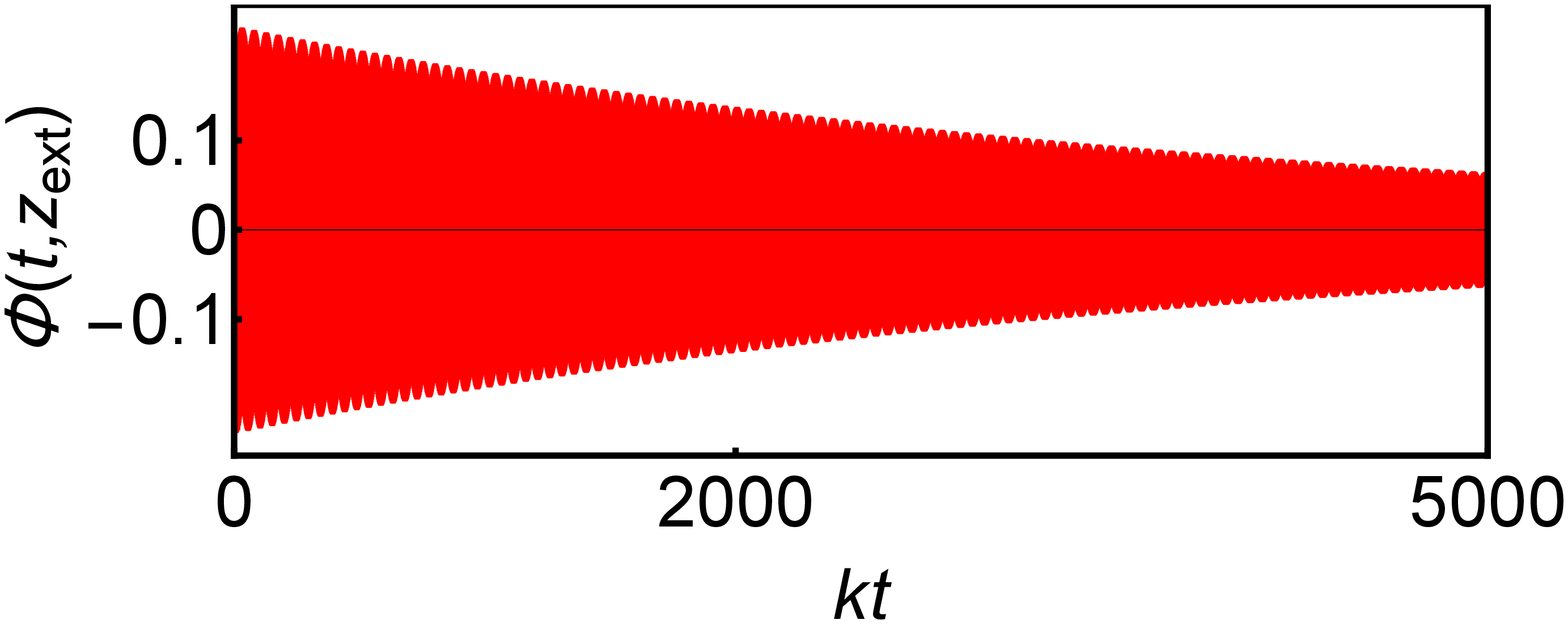}}	
	\subfigure[~$\alpha=7$]{\label{figfouriera7o1z1}
		\includegraphics[width=0.35\textwidth]{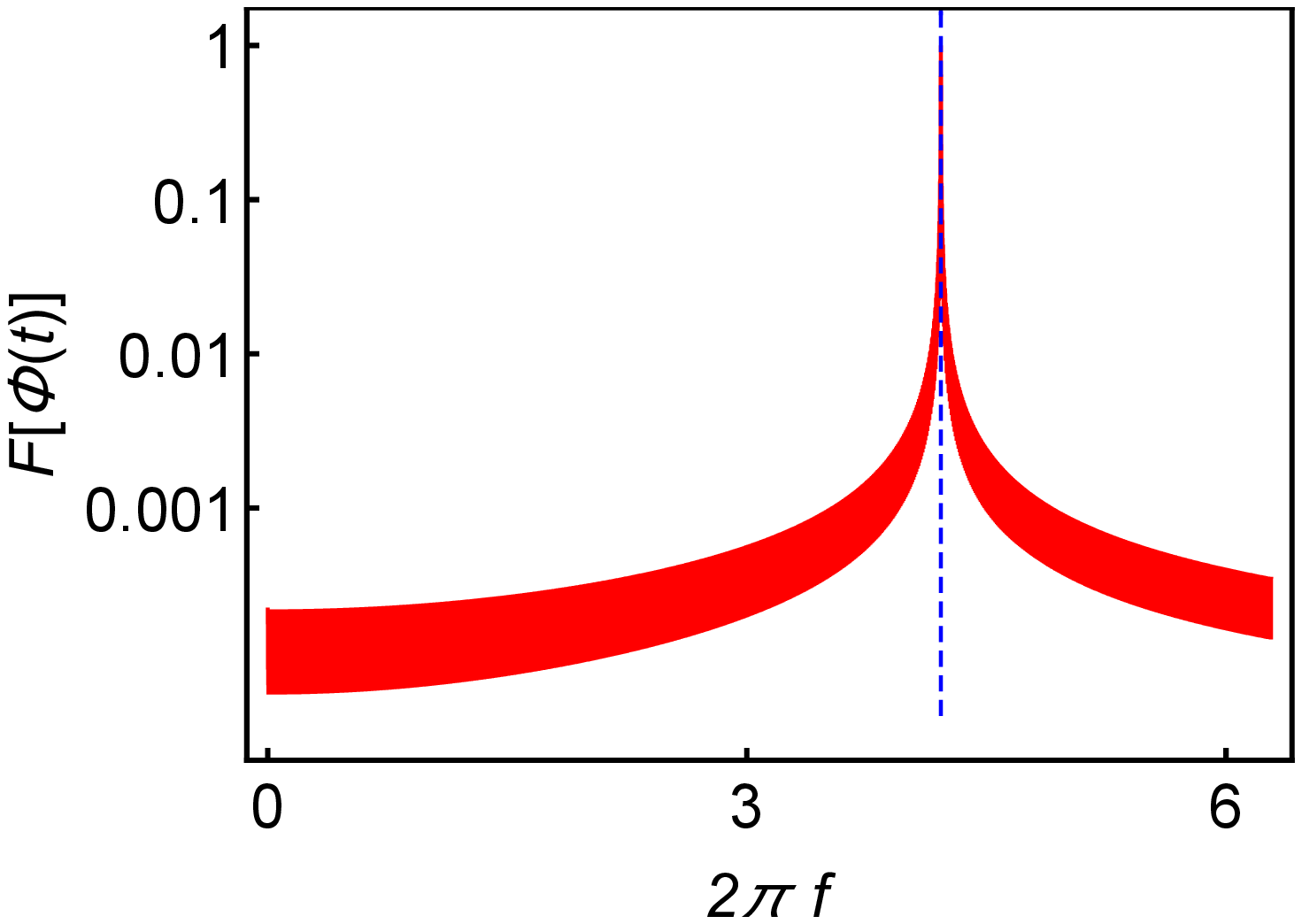}}	
	
	\subfigure[~$\alpha=9$]{\label{figfouriera9o1z1}
		\includegraphics[width=0.5\textwidth]{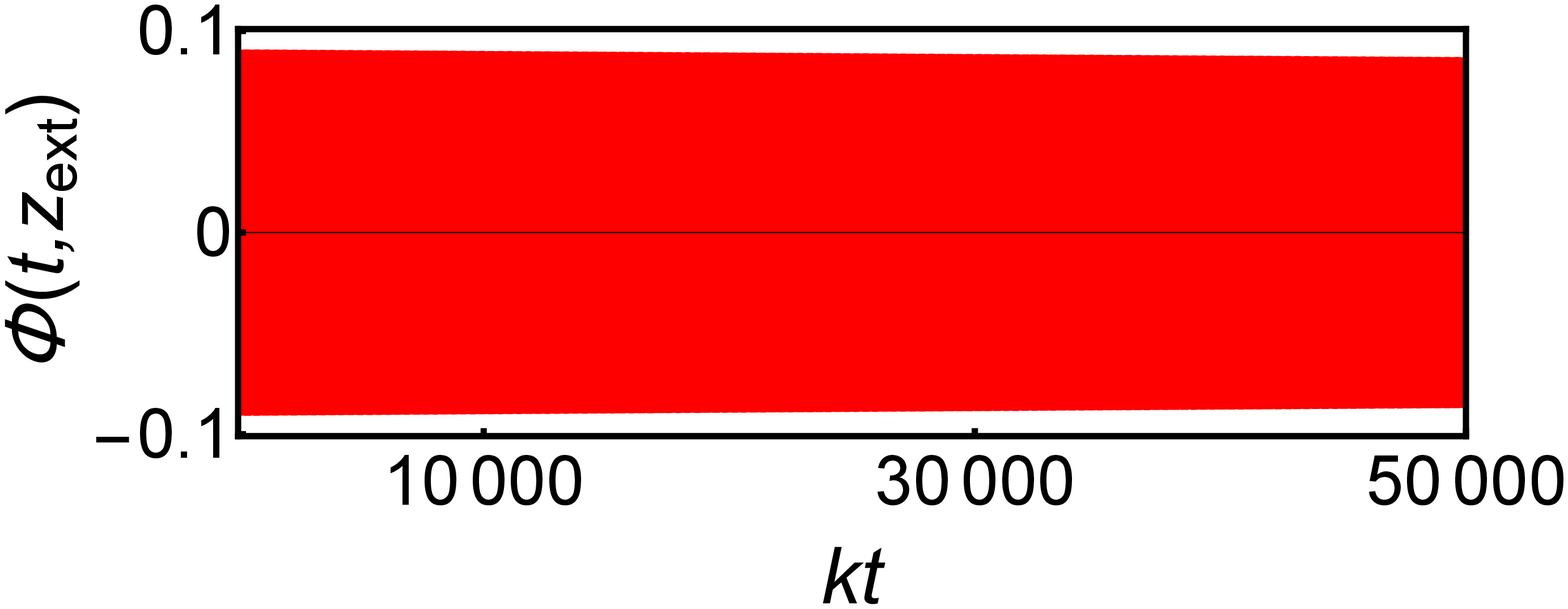}}	
	\subfigure[~$\alpha=9$]{\label{figa9o1z1}
		\includegraphics[width=0.35\textwidth]{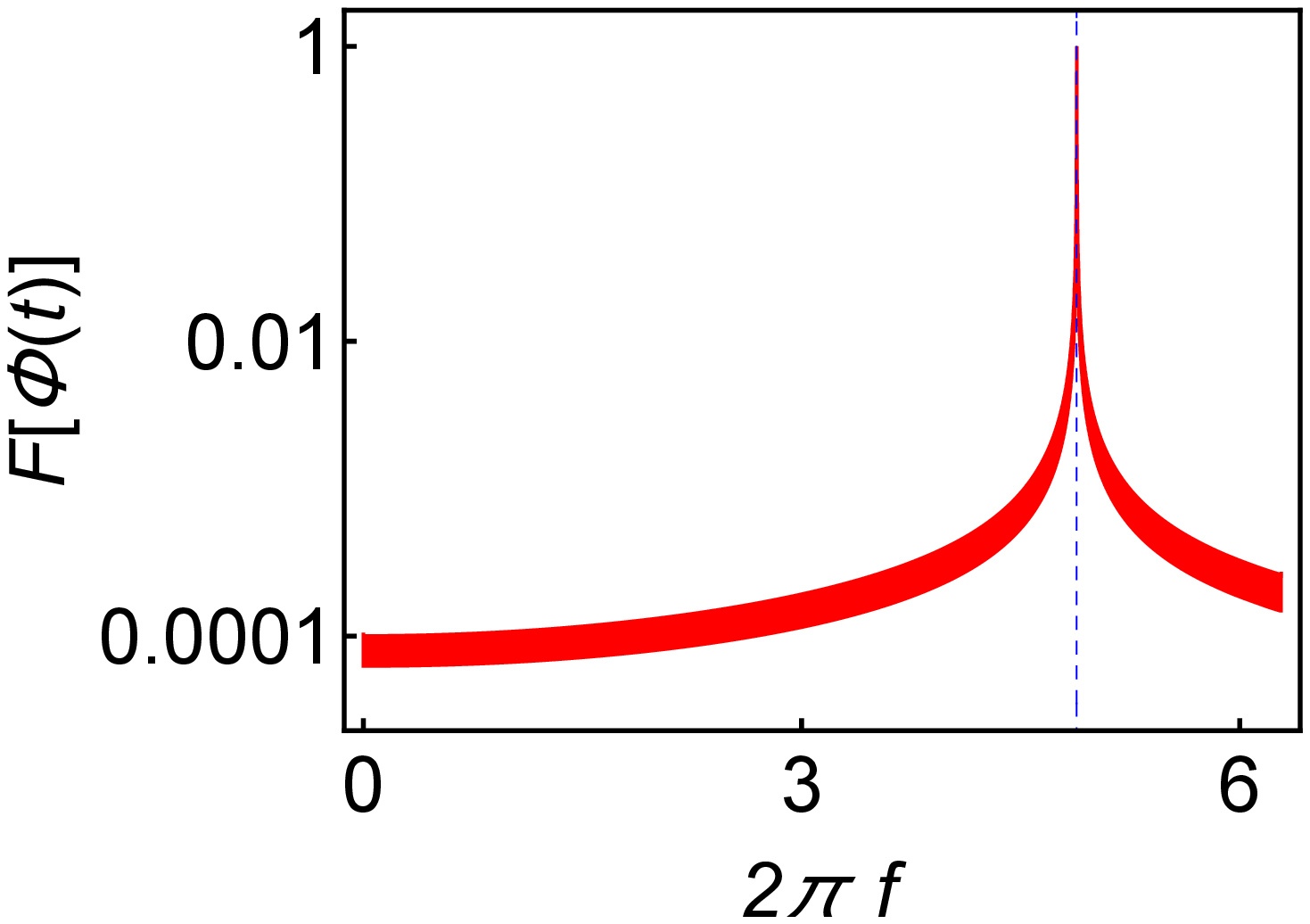}}	
	\caption{Left panel: Time evolution of the first gravitational resonance with $\alpha=5$ (top), $\alpha=7$ (middle), and $\alpha=9$ (bottom) at $kz_{\text{ext}}=1$. Right panel: The corresponding spectra in the frequency domain obtained by the discrete Fourier transformation. The dotted blue line corresponds to the KK mass of the first resonance.}\label{resonanceextfigandfourierfig}
\end{figure*}

As the effective potential is symmetric, the following boundary conditions can be used for numerical solution of the differential equation \eqref{Schrodingerlikeequation}:
\begin{subequations}
	\begin{eqnarray}
				\label{odd}
		\phi_{\rm{odd}}(0)\!\!&=&\!\!0, ~~~~\partial_{z}\phi_{\rm{odd}}(0)=1,\\
		\label{even}
		\phi_{\rm{even}}(0)\!\!&=&\!\!1, ~~~\partial_{z}\phi_{\rm{even}}(0)=0;
	\end{eqnarray}\label{EvenOddConditions}
\end{subequations}
where $\phi_{\rm{odd}}$ and $\phi_{\rm{even}}$ denote the odd and even modes of $\phi(z)$, respectively. Then we can numerically solve the gravitational resonances of the thick brane. We do not find any resonance for the case of $\alpha=1$. In fact, only for $\alpha\gtrsim2.8$, there are resonances. We plot the relative probability $P(m^{2})$ of gravitational KK modes for $\alpha=5, 7, 9$ in Fig.~\ref{figPalpha}. It can be seen that, the mass $m_{1}$ and relative probability $P(m_{1}^{2})$ of the first resonance increase with the parameter $\alpha$. Treating these gravitational resonances as the initial data, we can study the evolution of gravitational resonances on the brane~\cite{Tan:2022uex}. We use fourth-order finite differences in space and third-order Runge-Kutta integrator in time to solve the evolution equation \eqref{evolutionequation}. Since the thick brane is a dissipative system for the massive KK modes, i.e., the massive KK modes will escape to infinity of the extra dimension, we impose the maximally dissipative boundary condition~\cite{Megevand:2007uy}:
\begin{equation}
	\label{maximally dissipative boundary condition}
	\left\{
	\begin{aligned}
		\partial_t\phi&=&-\partial_z\phi, ~~~~~z\to\infty, \\
		\partial_t\phi&=&\partial_z\phi,  ~~~~z\to-\infty.
	\end{aligned}
	\right.
\end{equation}
Note that we only consider the case of $a=0$ in this paper. It means that the KK graviton propagates along the extra dimension at the speed of light, while the velocity component on the brane is zero. The boundary conditions of numerical evolution are easy to satisfy in this case. We analyze the results of the simulations by extracting a time series for the gravitational resonance amplitude at a fixed point $kz_{\text{ext}}=1$. In addition, we perform the discrete Fourier transformation to identify the oscillation frequencies of gravitational resonances. The discrete Fourier transform can be represented as follows:
\begin{eqnarray}
	F[\phi(t)](f):=|A\sum_{p}\phi(t_{p},z_{j})\text{exp}(-2\pi ift_{p} )|,\label{Fourier transform}
\end{eqnarray}
where $A$ is a normalization constant and $t_{p}$ represents discrete time values. The result can be seen from Fig.~\ref{resonanceextfigandfourierfig}. We can see that, the amplitude of resonance decreases with evolutionary time. This is reasonable, because the energy escape to infinity. The decay rate of resonance decreases with the parameter $\alpha$. This means that the lifetime of the resonance increases with $\alpha$. After the discrete Fourier transformation of the evolution of resonance, the KK mass of the resonance corresponds to the peak value of the spectra. This means that the KK mass of the resonance is the oscillation frequency of the resonance, that can be seen from Eq.~\eqref{momega} and $a=0$.

\begin{figure*}
	\subfigure[~$kz_{\text{ext}}=1$]{\label{figa12nrz1}
		\includegraphics[width=0.43\textwidth]{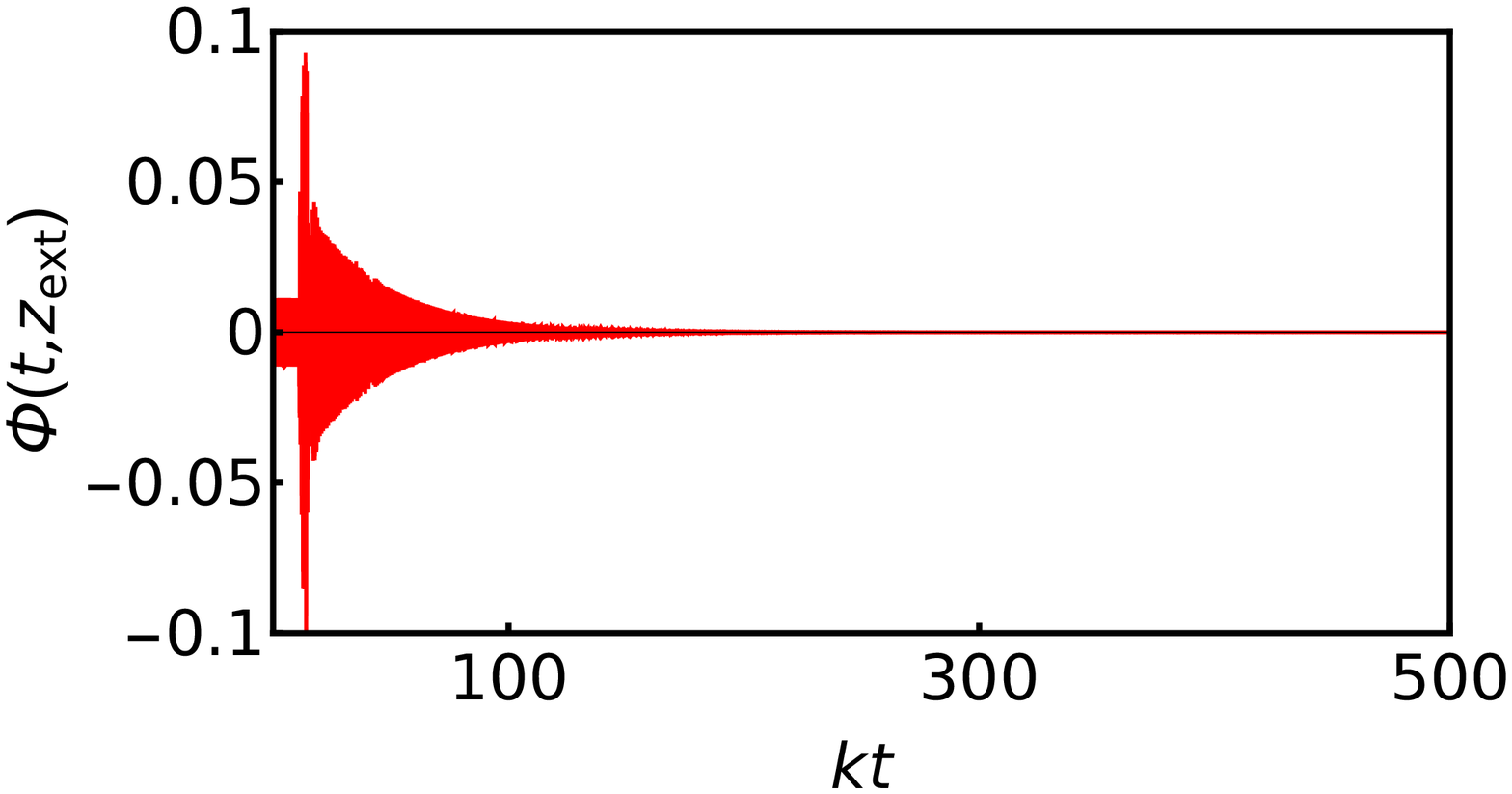}}
	\subfigure[~$kz_{\text{ext}}=1$]{\label{figfouriera12nrfig}
		\includegraphics[width=0.43\textwidth]{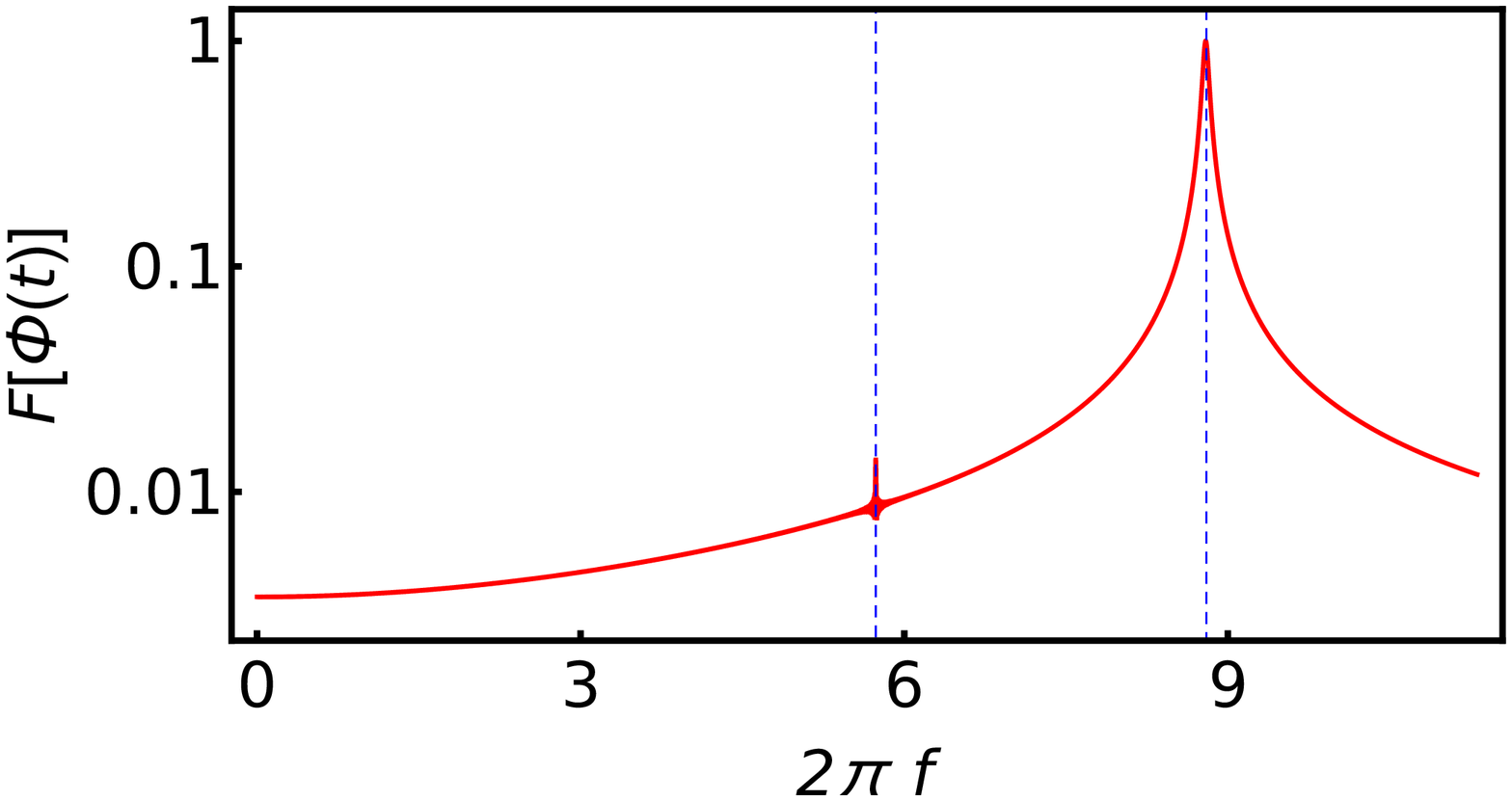}}		
	\caption{Left panel: Time evolution of the nonresonance with $\frac{m^{2}}{k^{2}}=70$ for $\alpha=12$. Right panel: Corresponding spectrum in the frequency domain obtained by the discrete Fourier transformation. The dotted blue lines correspond to the KK mass of the first two odd resonances.}\label{nonresonanceextfigandfourierfig}
\end{figure*}

\begin{figure}
	\subfigure[~$\alpha-\text{Re}(m_{1}/k)$]{\label{figRef}
		\includegraphics[width=0.22\textwidth]{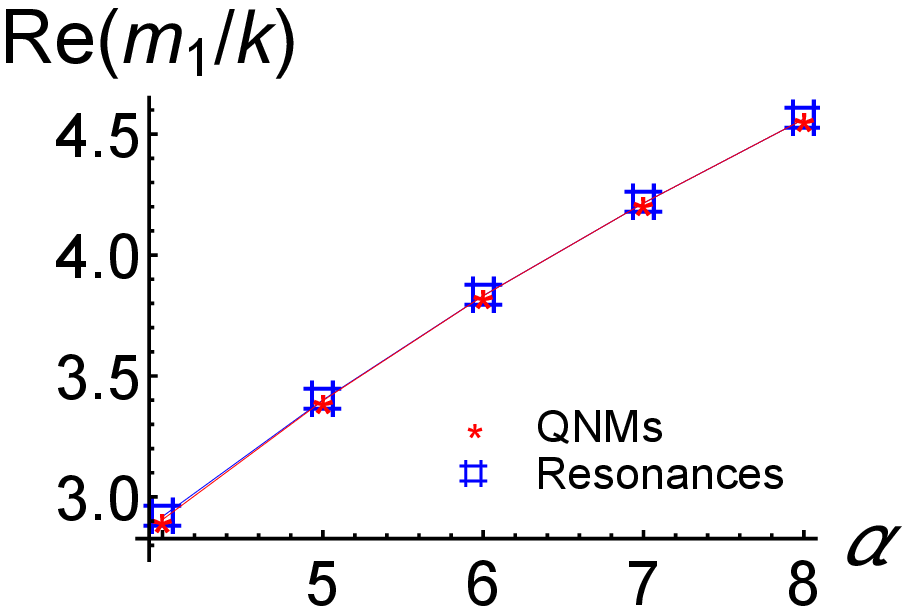}}
	\subfigure[~$\alpha-\text{Im}(m_{1}/k)$]{\label{figImf}
		\includegraphics[width=0.22\textwidth]{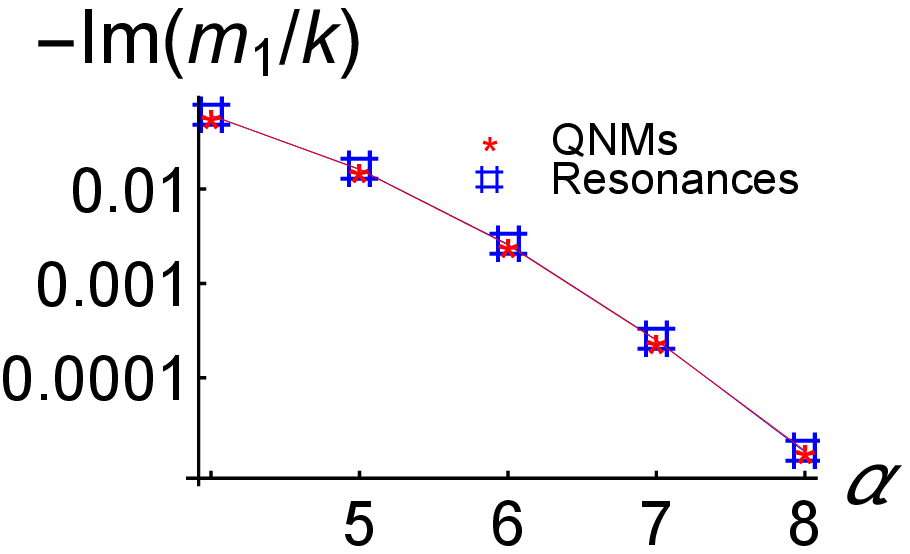}}		
	\caption{Left panel: The real parts of the first overtone QNFs which are obtained by the AIM and the KK mass of the first resonance. Right panel: The imaginary parts of the first overtone QNFs which are obtained by the AIM and the decay rate of the first resonance. Note that they are plotted on a logarithmic scale.}\label{figf}
\end{figure}

In addition, we perform the discrete Fourier transformation for a nonresonance, which can be seen from Fig.~\ref{nonresonanceextfigandfourierfig}. It can be seen that, there are two peaks in the frequency domain which correspond to the KK masses of the first two odd resonances. In other words, nonresonances evolve as combinations of resonances after some time. This seems to indicate that resonances are the characteristic modes of the thick brane. In fact, the oscillations of the resonances are equal (up to numerical error) to the real parts of the QNFs of the thick brane, while the decay rates of the resonances are equal to the imaginary parts of such ones, which can be seen from Fig.~\ref{figf}. Recall that the amplitudes of the resonances in the quasi-well are much larger than those outside the quasi-well. The long-lived QNMs are metastable states which quasi-localized on the brane. Thus we intuitively deem that these QNMs are localized near the thick brane (in the quasi-well). That is to say, the amplitudes of the long-lived QNMs in the quasi-well are also much larger than those outside ones, this corresponds to the situation of the resonances.

Finally, we investigate the half-life of the long-lived QNMs. For $\alpha=10$, the frequency of the first QNM is $m_{1}/k=5.18327-3.97621\times10^{-8}i$. If $k=10^{-3}$eV, the half-life $t_{1/2}$ of the first QNM is about $10^{-5}$s. This situation is quite different from the case of the RS-II brane. For the RS-II brane, the half-life of the first QNM is about $10^{-13}$s when $k=10^{-3}$eV \cite{Seahra:2005wk}. From Fig.~\ref{alpham} we can see that, the imaginary part of the first QNF decreases with $\alpha$. This means that the lifetime of the first QNM increases with $\alpha$. Thus, for a large enough $\alpha$, the first QNM has a very long lifetime. We expect that the long-lived modes will be detected in the future~\cite{Bian:2021ini,Guo:2022sts}. In addition, recent work by Teukolsky et al. on black hole QNMs indicates that overtone modes of black holes are more significant than previously thought~\cite{Giesler:2019uxc}. These overtone modes can dominate at the early stage of the ringdown. Consequently, the nature of overtone modes in thick branes is also worthy of further investigation. We will explore the characteristics of overtone modes in thick branes in our future research.

\section{Conclusion and discussion}
\label{Conclusion}
In this paper, we studied the QNMs and the resonances of the thick brane model. We found that the oscillations of the resonances equal to the real parts of the QNFs of the thick brane, while the decay rates of the resonances equal to the imaginary parts of those QNFs. These QNMs can exist for a very long time on the brane, perhaps even on the cosmological time scale. They might be viewed as a candidate for dark matter and might be detected as a stochastic gravitational wave background~\cite{Chen:2021ntg}.

At the beginning of this paper, we reviewed the thick brane model in the five-dimensional spacetime. By studying the linear transverse-traceless tensor fluctuation~(\ref{perturbed metric}), we obtained the evolution equation (\ref{evolutionequation}) and the Schr\"odinger-like equation~(\ref{Schrodingerlikeequation}). Then we used the asymptotic iteration and shooting methods to solve the QNFs of the thick brane. From Table~\ref{tab1}, we can see that, the results of the two methods are consistent with each other in the low overtones. By investigating the effect of the parameter $\alpha$ on the QNMs of the brane, we found that the real parts of the first two QNFs increase with $\alpha$, while the imaginary parts of the first three QNFs decrease with $\alpha$, which can be seen from Fig.~\ref{alpham}. Since the imaginary parts of the QNFs correspond to the damping rates of the QNMs, these KK modes could become long-lived modes when the parameter $\alpha$ is large enough.

On the other hand, the resonances will appear when $\alpha$ is large. Thus we suspect that the long-lived QNMs are related to the resonances. To verify this, we investigated the resonances of this brane by the relative probability method. Then, we investigated the evolution of these resonances, which can be seen from Figs.~\ref{figPalpha},~\ref{resonanceextfigandfourierfig}, and~\ref{figf}. The results show that the oscillations of the resonances equal to the real parts of the QNFs of the thick brane, while the damping rates of the resonances equal to the imaginary parts of the QNFs. Finally, we investigated the half-life of the long-lived QNMs. For a very large $\alpha$, these QNMs could exist for a very long time on the brane, perhaps even on the cosmological time scale.

There is a lot to be improved in this paper. For example, the QNMs and the evolution of other test fields could be investigated. The effect of these long-lived KK modes on the stochastic gravitational wave background is also worth investigating.

\section*{Acknowledgements}
This work was supported by the National Key Research and Development Program of China (Grant No. 2020YFC2201503), the National Natural Science Foundation of China (Grants No. 12347111,
No. 12205129, No.~11875151, No.~12105126, No. 12147166, and No.~12247101), the 111 Project under (Grant No. B20063), the China Postdoctoral Science Foundation (Grant No. 2021M701529 and
No. 2023M741148), the Major Science and Technology Projects of Gansu Province, and ``Lanzhou City's scientific research funding subsidy to Lanzhou University".

\end{document}